\documentclass[12pt,a4paper]{article}

\usepackage{ifthen} 
\usepackage{subcaption}
\newboolean{pdflatex}
\setboolean{pdflatex}{true} 

\newboolean{articletitles}
\setboolean{articletitles}{true} 

\newboolean{uprightparticles}
\setboolean{uprightparticles}{false} 

\def\paperauthors{LHCb collaboration} 
\def\paperasciititle{Measurement of the Z boson production cross-section in proton-lead collisions at sqrt(s_NN)=8.16 TeV} 
\def\papertitle{Measurement of the $Z$ boson production cross-section in proton-lead collisions at $\sqrt{s_\mathrm{NN}}=8.16\,\mathrm{TeV}$} 
\def\paperkeywords{{High Energy Physics}, {LHCb}} 
\def\papercopyright{\the\year\ CERN for the benefit of the LHCb collaboration} 
\def\paperlicence{CC BY 4.0 licence}
\def\paperlicenceurl{https://creativecommons.org/licenses/by/4.0/}

\usepackage[top=1in, bottom=1.25in, left=1in, right=1in]{geometry}

\usepackage{microtype}
\usepackage{lineno}  
\usepackage{xspace} 
\usepackage{caption} 

\usepackage{graphicx}  
\usepackage{color}
\usepackage{colortbl}
\graphicspath{{./figs/}} 

\usepackage{amsmath} 
\usepackage{amssymb}
\usepackage{amsfonts}
\usepackage{upgreek} 

\newcommand*\patchAmsMathEnvironmentForLineno[1]{%
\expandafter\let\csname old#1\expandafter\endcsname\csname #1\endcsname
\expandafter\let\csname oldend#1\expandafter\endcsname\csname
end#1\endcsname
 \renewenvironment{#1}%
   {\linenomath\csname old#1\endcsname}%
   {\csname oldend#1\endcsname\endlinenomath}%
}
\newcommand*\patchBothAmsMathEnvironmentsForLineno[1]{%
  \patchAmsMathEnvironmentForLineno{#1}%
  \patchAmsMathEnvironmentForLineno{#1*}%
}
\AtBeginDocument{%
\patchBothAmsMathEnvironmentsForLineno{equation}%
\patchBothAmsMathEnvironmentsForLineno{align}%
\patchBothAmsMathEnvironmentsForLineno{flalign}%
\patchBothAmsMathEnvironmentsForLineno{alignat}%
\patchBothAmsMathEnvironmentsForLineno{gather}%
\patchBothAmsMathEnvironmentsForLineno{multline}%
\patchBothAmsMathEnvironmentsForLineno{eqnarray}%
}

\usepackage{hyperxmp}

\usepackage[pdftex,
            pdfauthor={\paperauthors},
            pdftitle={\paperasciititle},
            pdfkeywords={\paperkeywords},
            pdfcopyright={Copyright (C) \papercopyright},
            pdflicenseurl={\paperlicenceurl}]{hyperref}

\usepackage[colorinlistoftodos,textsize=scriptsize]{todonotes}

\usepackage[bottom,flushmargin,hang]{footmisc}

\usepackage[all]{hypcap} 

\usepackage{xspace} 
\usepackage{upgreek}

\def\lhcb   {\mbox{LHCb}\xspace}
\def\atlas  {\mbox{ATLAS}\xspace}
\def\cms    {\mbox{CMS}\xspace}
\def\alice  {\mbox{ALICE}\xspace}

\def\lhc    {\mbox{LHC}\xspace}

\def\MagUp {\mbox{\em Mag\kern -0.05em Up}\xspace}

\ifthenelse{\boolean{uprightparticles}}%
{

 \def\Pmu         {\ensuremath{\upmu}\xspace}

 \def\Ptau        {\ensuremath{\uptau}\xspace}

 \def\PDelta      {\ensuremath{\Delta}\xspace}                 
 \def\PXi         {\ensuremath{\Xi}\xspace}                 
 \def\PLambda     {\ensuremath{\Lambda}\xspace}                 
 \def\PSigma      {\ensuremath{\Sigma}\xspace}                 
 \def\POmega      {\ensuremath{\Omega}\xspace}                 
 \def\PUpsilon    {\ensuremath{\Upsilon}\xspace}

 \def\PB      {\ensuremath{\mathrm{B}}\xspace}                 
                  
 \def\PD      {\ensuremath{\mathrm{D}}\xspace}

 \def\PK      {\ensuremath{\mathrm{K}}\xspace}

 \def\PW      {\ensuremath{\mathrm{W}}\xspace}

 \def\PZ      {\ensuremath{\mathrm{Z}}\xspace}

 \def\Pi      {\ensuremath{\mathrm{i}}\xspace}

 \def\Pp      {\ensuremath{\mathrm{p}}\xspace}

 \def\Ps      {\ensuremath{\mathrm{s}}\xspace}                 
 \def\Pt      {\ensuremath{\mathrm{t}}\xspace}

 \def\thebaroffset{0.0em}
}
{

 \def\Pmu         {\ensuremath{\mu}\xspace}

 \def\Ptau        {\ensuremath{\tau}\xspace}

 \mathchardef\PDelta="7101
 \mathchardef\PXi="7104
 \mathchardef\PLambda="7103
 \mathchardef\PSigma="7106
 \mathchardef\POmega="710A
 \mathchardef\PUpsilon="7107
                  
 \def\PB      {\ensuremath{B}\xspace}                 
                  
 \def\PD      {\ensuremath{D}\xspace}

 \def\PK      {\ensuremath{K}\xspace}

 \def\PW      {\ensuremath{W}\xspace}

 \def\PZ      {\ensuremath{Z}\xspace}

 \def\Pi      {\ensuremath{i}\xspace}

 \def\Pp      {\ensuremath{p}\xspace}

 \def\Ps      {\ensuremath{s}\xspace}                 
 \def\Pt      {\ensuremath{t}\xspace}

 \def\thebaroffset{0.18em}
}
\newcommand{\offsetoverline}[2][\thebaroffset]{\kern #1\overline{\kern -#1 #2}}%

\makeatletter
\ifcase \@ptsize \relax
  \newcommand{\miniscule}{\@setfontsize\miniscule{4}{5}}
\or
  \newcommand{\miniscule}{\@setfontsize\miniscule{5}{6}}
\or
  \newcommand{\miniscule}{\@setfontsize\miniscule{5}{6}}
\fi
\makeatother

\DeclareRobustCommand{\optbar}[1]{\shortstack{{\miniscule (\rule[.5ex]{1.25em}{.18mm})}
  \\ [-.7ex] $#1$}}

\def\mumu       {{\ensuremath{\Pmu^+\Pmu^-}}\xspace}

\def\tautau     {{\ensuremath{\Ptau^+\Ptau^-}}\xspace}

\def\Wp     {{\ensuremath{\PW^+}}\xspace}
\def\Wm     {{\ensuremath{\PW^-}}\xspace}

\def\Z      {{\ensuremath{\PZ}}\xspace}

\def\squark    {{\ensuremath{\Ps}}\xspace}

\def\tquark    {{\ensuremath{\Pt}}\xspace}
\def\tquarkbar {{\ensuremath{\overline \tquark}}\xspace}
\def\ttbar     {{\ensuremath{\tquark\tquarkbar}}\xspace}

\def\KorKbar {\kern \thebaroffset\optbar{\kern -\thebaroffset \PK}{}\xspace}

\def\D       {{\ensuremath{\PD}}\xspace}

\def\DorDbar {\kern \thebaroffset\optbar{\kern -\thebaroffset \PD}\xspace}

\def\Dp      {{\ensuremath{\D^+}}\xspace}
\def\Dm      {{\ensuremath{\D^-}}\xspace}

\def\DpDm    {\ensuremath{\Dp {\kern -0.16em \Dm}}\xspace}

\def\B       {{\ensuremath{\PB}}\xspace}

\def\BorBbar {\kern \thebaroffset\optbar{\kern -\thebaroffset \PB}\xspace}

\def\Bd      {{\ensuremath{\B^0}}\xspace}

\def\BdorBdbar {\kern \thebaroffset\optbar{\kern -\thebaroffset \Bd}\xspace}

\def\Bs      {{\ensuremath{\B^0_\squark}}\xspace}

\def\BsorBsbar {\kern \thebaroffset\optbar{\kern -\thebaroffset \Bs}\xspace}

\def\Y#1S{\ensuremath{\PUpsilon{(#1S)}}\xspace}

\def\proton      {{\ensuremath{\Pp}}\xspace}

\def\LorLbar     {\kern \thebaroffset\optbar{\kern -\thebaroffset \PLambda}\xspace}

\newcommand{\decay}[2]{\ensuremath{#1\!\to #2}\xspace} 

\def\to                 {\ensuremath{\rightarrow}\xspace}

\def\AT#1     {\ensuremath{A_{\mathrm{T}}^{#1}}\xspace}           

\def\C#1      {\ensuremath{\mathcal{C}_{#1}}\xspace}                       
\def\Cp#1     {\ensuremath{\mathcal{C}_{#1}^{'}}\xspace}                    
\def\Ceff#1   {\ensuremath{\mathcal{C}_{#1}^{\mathrm{(eff)}}}\xspace}        
\def\Cpeff#1  {\ensuremath{\mathcal{C}_{#1}^{'\mathrm{(eff)}}}\xspace}       
\def\Ope#1    {\ensuremath{\mathcal{O}_{#1}}\xspace}                       
\def\Opep#1   {\ensuremath{\mathcal{O}_{#1}^{'}}\xspace}                    

\newcommand{\nospaceunit}[1]{\ensuremath{\text{#1}}}       
\newcommand{\aunit}[1]{\ensuremath{\text{\,#1}}}       

\newcommand{\tev}{\aunit{Te\kern -0.1em V}\xspace}
\newcommand{\gev}{\aunit{Ge\kern -0.1em V}\xspace}
\newcommand{\mev}{\aunit{Me\kern -0.1em V}\xspace}
\newcommand{\kev}{\aunit{ke\kern -0.1em V}\xspace}
\newcommand{\ev}{\aunit{e\kern -0.1em V}\xspace}
 
\newcommand{\mevc}{\ensuremath{\aunit{Me\kern -0.1em V\!/}c}\xspace}
\newcommand{\gevc}{\ensuremath{\aunit{Ge\kern -0.1em V\!/}c}\xspace}
\newcommand{\mevcc}{\ensuremath{\aunit{Me\kern -0.1em V\!/}c^2}\xspace}
\newcommand{\gevcc}{\ensuremath{\aunit{Ge\kern -0.1em V\!/}c^2}\xspace}

\def\mum  {\ensuremath{\,\upmu\nospaceunit{m}}\xspace}

\def\nb {\aunit{nb}\xspace}
\def\invnb {\ensuremath{\nb^{-1}}\xspace}
\def\pb {\aunit{pb}\xspace}

\def\fb   {\ensuremath{\aunit{fb}}\xspace}
\def\invfb   {\ensuremath{\fb^{-1}}\xspace}

\newcommand{\chisqndf}{\ensuremath{\chi^2/\mathrm{ndf}}\xspace}
\newcommand{\chisqip}{\ensuremath{\chi^2_{\text{IP}}}\xspace}

\def\deriv {\ensuremath{\mathrm{d}}}

\def\gsim{{~\raise.15em\hbox{$>$}\kern-.85em
          \lower.35em\hbox{$\sim$}~}\xspace}
\def\lsim{{~\raise.15em\hbox{$<$}\kern-.85em
          \lower.35em\hbox{$\sim$}~}\xspace}

\def\sqs   {\ensuremath{\protect\sqrt{s}}\xspace}
\def\sqsnn {\ensuremath{\protect\sqrt{s_{\scriptscriptstyle\text{NN}}}}\xspace}
\def\pt         {\ensuremath{p_{\mathrm{T}}}\xspace}

\def\ptot       {\ensuremath{p}\xspace}

\newcommand{\lum} {\ensuremath{\mathcal{L}}\xspace}

\def\evtgen     {\mbox{\textsc{EvtGen}}\xspace}

\def\geant      {\mbox{\textsc{Geant4}}\xspace}

\def\photos     {\mbox{\textsc{Photos}}\xspace}
\def\powheg     {\mbox{\textsc{PowhegBox}}\xspace}
\def\pythia     {\mbox{\textsc{Pythia}}\xspace}

\def\tell1  {TELL1\xspace}
\def\ukl1   {UKL1\xspace}

\newcommand{\eg}{\mbox{\itshape e.g.}\xspace}
\newcommand{\ie}{\mbox{\itshape i.e.}\xspace}

\newcommand{\lhcborcid}[1]{\href{https://orcid.org/#1}{\hspace*{0.1em}\raisebox{-0.45ex}{\includegraphics[width=1em]{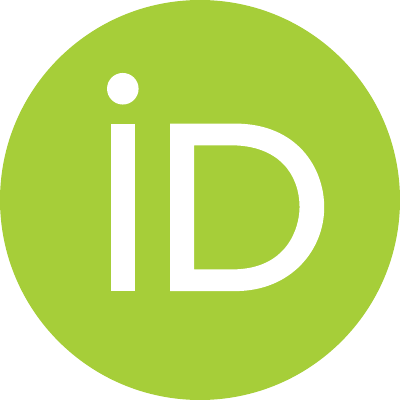}}}}

\newcommand{\pz}{\phantom{0}}

\def\gevcsquare    {\ensuremath{{\,(\mathrm{Ge\kern -0.1em V\!/}c)^2}}\xspace}
\def\mygevc        {\ensuremath{{\mathrm{Ge\kern -0.1em V\!/}c}}\xspace}
\def\unitgev   {\ensuremath{\mathrm{[Ge\kern -0.1em V]}}\xspace}

\def\pPb   {\ensuremath{\proton\mathrm{Pb}}\xspace}
\def\Pbp   {\ensuremath{\mathrm{Pb}\proton}\xspace}

\def\For    {Forward\xspace}
\def\Back   {Backward\xspace}
\def\forw   {forward\xspace}
\def\backw  {backward\xspace}

\def\rfb   {\ensuremath{R_\mathrm{FB}}\xspace}
\def\rpa   {\ensuremath{R_{\proton \mathrm{Pb}}}\xspace}
\def\kfb   {\ensuremath{k_\mathrm{FB}}\xspace}
\def\kpa   {\ensuremath{k_{\proton \mathrm{Pb}}}\xspace}

\def\Zmumu   {\ensuremath{\Z\to\mumu}\xspace}

\def\effreco {\ensuremath{\epsilon^{\mathrm{reco\&sel}}}\xspace}
\def\effmuid {\ensuremath{\epsilon^{\mathrm{muon{\text{-}}id}}}\xspace}
\def\efftrig {\ensuremath{\epsilon^{\mathrm{trig}}}\xspace}

\def\fsrcorr {\ensuremath{f_{\mathrm{FSR}}}\xspace}
\def\ncand   {\ensuremath{N_{\mathrm{cand}}}\xspace}

\def\phistar {\ensuremath{\phi^{*}}\xspace}
\def\zrap {\ensuremath{y_{\Z}}\xspace}
\def\zrapstar {\ensuremath{y^{*}_{\Z}}\xspace}
\def\ZpT     {\ensuremath{p_{\mathrm{T}}^{\Z}}\xspace}
\def\Iso     {\ensuremath{I_{\mathrm{T}}^{\mathrm{charged}}}\xspace}
\def\murapstar {\ensuremath{y^{*}_{\mu}}\xspace}
\def\fidxsec {\ensuremath{\sigma^{\mathrm{fid}}}\xspace}

\def\mupt    {\ensuremath{p_{\mathrm{T}}^{\mu}}\xspace}
\def\meta    {\ensuremath{{\eta}^{\mu}}\xspace}

\def\lhapdf     {\mbox{\textsc{Lhapdf}}\xspace}

\usepackage{cite} 
\usepackage{mciteplus}

\usepackage{longtable} 
\usepackage{rotating}

\begin{document}

\renewcommand{\thefootnote}{\fnsymbol{footnote}}
\setcounter{footnote}{1}


\begin{titlepage}
\pagenumbering{roman}

\vspace*{-1.5cm}
\centerline{\large EUROPEAN ORGANIZATION FOR NUCLEAR RESEARCH (CERN)}
\vspace*{1.5cm}
\noindent
\begin{tabular*}{\linewidth}{lc@{\extracolsep{\fill}}r@{\extracolsep{0pt}}}
\ifthenelse{\boolean{pdflatex}}
{\vspace*{-1.5cm}\mbox{\!\!\!\includegraphics[width=.14\textwidth]{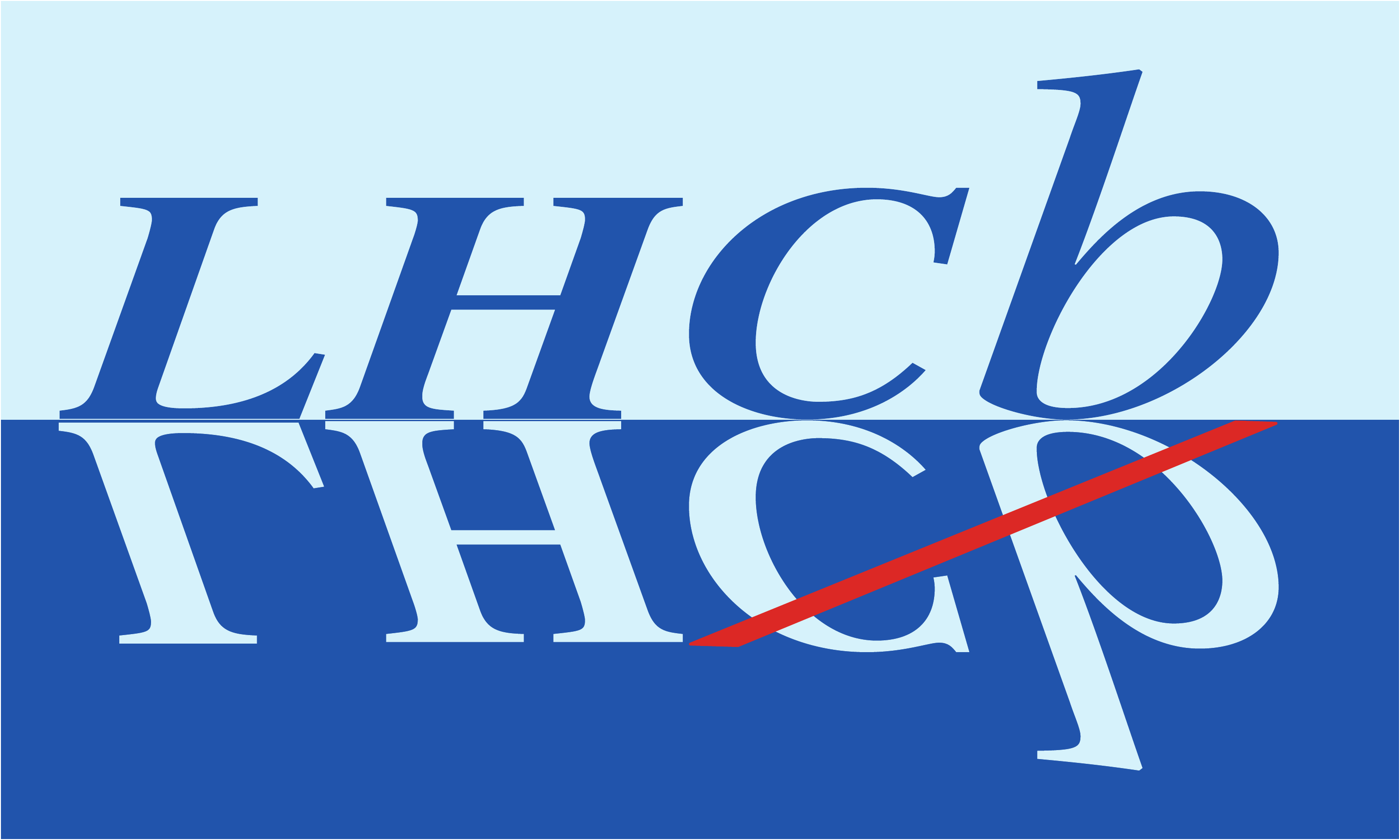}} & &}%
{\vspace*{-1.2cm}\mbox{\!\!\!\includegraphics[width=.12\textwidth]{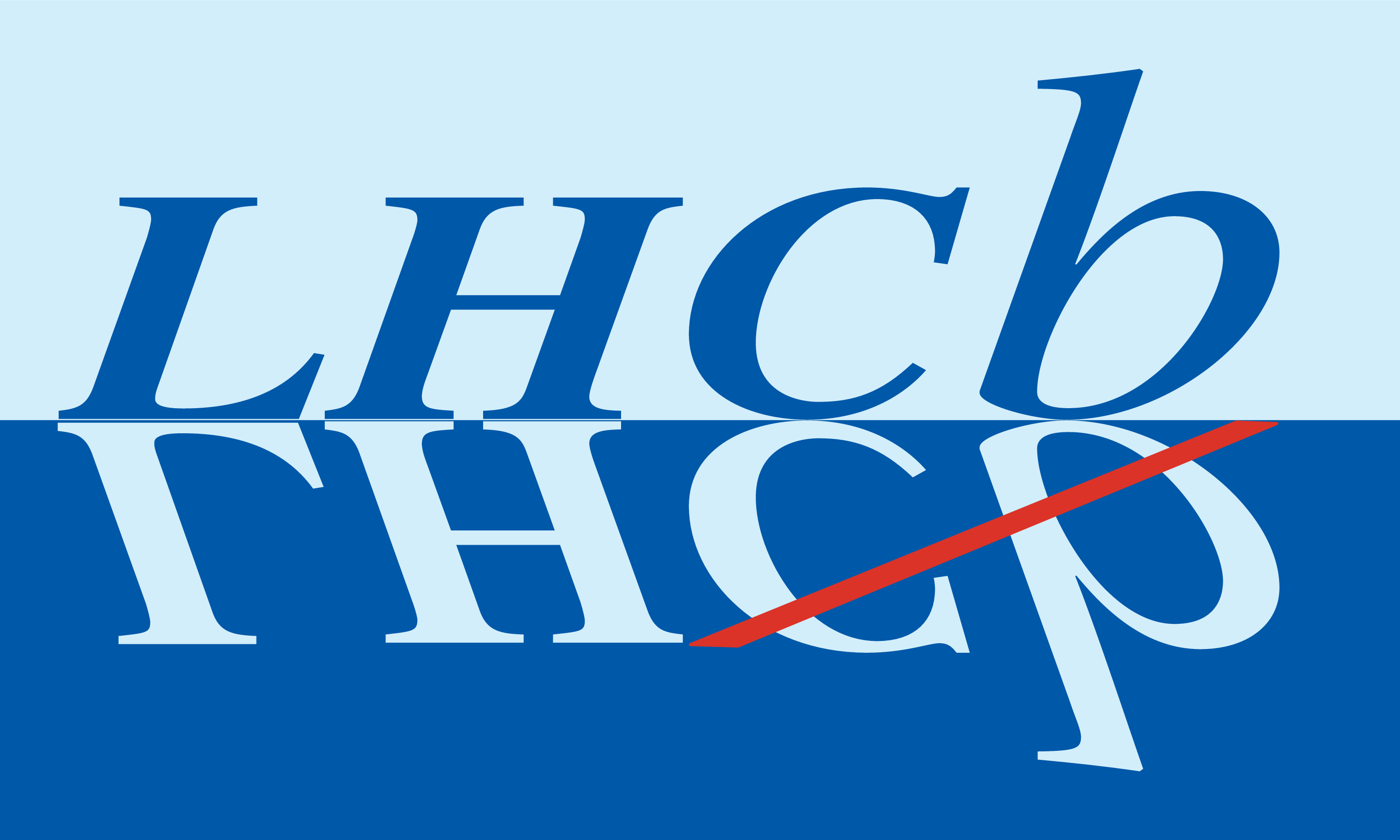}} & &}%
\\
 & & CERN-EP-2022-089 \\  
 & & LHCb-PAPER-2022-009 \\  
 & & 8 June 2023 \\ 
 & & \\
\end{tabular*}

\vspace*{2.0cm}

{\normalfont\bfseries\boldmath\huge
\begin{center}
  \papertitle 
\end{center}
}

\vspace*{2.0cm}

\begin{center}

\paperauthors\footnote{Authors are listed at the end of this paper.}
\end{center}

\vspace{\fill}

\begin{abstract}
  \noindent
This article presents the first measurement of the differential $Z$-boson production cross-section in the forward region using proton-lead collisions with the LHCb detector. The dataset was collected at a nucleon-nucleon centre-of-mass energy of $\sqrt{s_\mathrm{NN}}=8.16\,\mathrm{TeV}$ in 2016, corresponding to an integrated luminosity of $30.8\,\mathrm{nb}^{-1}$. The forward-backward ratio and the nuclear modification factors are measured together with the differential cross-section as functions of the $Z$ boson rapidity in the centre-of-mass frame, the transverse momentum of the $Z$ boson and a geometric variable $\phi^{*}$. The results are in good agreement with the predictions from nuclear parton distribution functions, providing strong constraining power at small \mbox{Bjorken-$x$}.

\end{abstract}

\vspace*{2.0cm}

\begin{center}
  Published in JHEP 06 (2023) 022
\end{center}

\vspace{\fill}

{\footnotesize 
\centerline{\copyright~\papercopyright. \href{\paperlicenceurl}{\paperlicence}.}}
\vspace*{2mm}

\end{titlepage}


\newpage
\setcounter{page}{2}
\mbox{~}


\renewcommand{\thefootnote}{\arabic{footnote}}
\setcounter{footnote}{0}

\cleardoublepage


\pagestyle{plain} 
\setcounter{page}{1}
\pagenumbering{arabic}

\section{Introduction}
\label{sec:Introduction}

Inclusive \Z-boson\footnote{In this paper, the label \Z boson is defined
to include contributions from virtual photons.}
production at hadron colliders~\cite{Drell:1970wh}
is an important benchmark process to test quantum chromodynamics (QCD),
and can be factorised~\cite{Collins:1989gx} as a product
of the hard scattering and the initial 
state of the collision.
State-of-the-art next-to-next-to-leading order (NNLO) calculations in perturbative QCD (pQCD), together with
next-to-next-to-leading logarithm (NNLL) resummation~\cite{Ladinsky:1993zn,Balazs:1995nz, Balazs:1997sk,Balazs:1997xd,Landry:2002ix,
Konychev:2005iy,Camarda:2019zyx,Camarda:2021ict} and 
next-to-leading order (NLO) 
electroweak (EW) corrections~\cite{Hamberg:1990np,
Catani:2009sm, Melnikov:2006kv}, precisely describe the hard process, while the non-perturbative initial-state can be be modelled through 
parton distribution functions (PDF)~\cite{Rojo:2015acz, Butterworth:2015oua} or
nuclear PDFs (nPDF)~\cite{Kusina:2016fxy, 
Eskola:2016oht, deFlorian:2003qf, Hirai:2007sx, AtashbarTehrani:2012xh, Khanpour:2016pph}.
As a result, \Z-boson production
at the Large Hadron Collider (\lhc) carries valuable information in 
constraining the PDFs and nPDFs. 
Weakly coupled \Z bosons and their leptonic decay final states have a negligible 
interaction with the nuclear medium and
can be used as clean probes of nuclear-matter effects on the initial-state. 
The production of \Z bosons is sensitive to only the initial-state, while hadronic probes are sensitive to both initial- and final-state nuclear matter effects. Therefore, together with hadronic probes \Z-boson production can differentiate between effects 
of the initial- and final-state.
Studies~\cite{Kusina:2016fxy} also show that \Z-boson production in proton-lead (\pPb) collisions
at the \lhc is sensitive to heavier quark flavours. 
Improved information on the nuclear corrections is helpful to reduce proton PDF uncertainties and
essential for distinguishing the distributions for the different parton flavours.
Moreover, \Z-boson production is an ideal process to probe transverse-momentum-dependent PDFs (TMDs)~\cite{Brodsky:2002cx, 
Collins:2002kn, Boer:2003cm,Belitsky:2002sm} when its transverse
momentum is below $\sim 10\gev$, where non-perturbative QCD effects start to dominate~\cite{Peng:2014hta}.

The \lhcb experiment published for the first time the inclusive \Z production result 
in the forward region in \pPb collisions at a nucleon-nucleon centre-of-mass energy \mbox{$\sqsnn=5.02\tev$~\cite{LHCb-PAPER-2014-022}.} 
The unique forward geometry coverage allows the \lhcb detector to probe
nPDFs at very small Bjorken-$x$ ($10^{-4}<x<10^{-3}$).
This measurement was followed by the inclusive and differential results in the central region from the \cms and
\atlas experiments~\cite{Khachatryan:2015pzs,
Aad:2015gta}
and in the forward region from the \alice experiment~\cite{Alice:2016wka} at \mbox{$\sqsnn=5.02\tev$}. 
Inclusive and differential results are also measured at $\sqsnn=8.16\tev$ 
by the \cms collaboration in the
central region~\cite{CMS:2021ynu} and by the \alice collaboration in the forward region~\cite{ALICE:2020jff}.
These results are in good agreement with NLO pQCD predictions calculated with
commonly used nPDFs such as EPPS09~\cite{Eskola:2016oht} and nCTEQ15~\cite{Kovarik:2015cma}.

This article presents the first differential measurement of the production cross-section,
the forward-backward ratio of the production cross-sections and the nuclear modification factors of \decay{\Z}{\mumu} production with the \lhcb experiment
using \pPb collisions.
The dataset was collected at $\sqsnn=8.16\tev$ in 2016.
The inclusive cross-sections in 
the proton-lead (forward)
and lead-proton (backward) collisions are measured,
together with the differential cross-sections
as a function of the rapidity of the \Z boson in the centre-of-mass frame (\zrapstar),
the transverse momentum (\ZpT) and an angular variable \phistar~\cite{Vesterinen:2008hx}.
The ratio between the cross-sections in forward and backward collisions in a common rapidity
range (\rfb) and the nuclear modification factors (\rpa) with respect to $pp$ collisions are measured, both inclusively and differentially as functions of the above three variables.

\section{Detector, datasets, and theoretical predictions}
\label{sec:Detector}

The \lhcb detector~\cite{LHCb-DP-2008-001, LHCb-DP-2014-002} 
is a single-arm forward
spectrometer covering the 
pseudorapidity ($\eta$) range from 2 to 5.
The detector includes a high-precision tracking system
consisting of a silicon-strip vertex detector surrounding the $pp$
interaction region, a large-area silicon-strip detector located
upstream of a dipole magnet with a bending power of about
$4{\mathrm{\,Tm}}$, and three stations of silicon-strip detectors and straw
drift tubes placed downstream of the magnet.
The tracking system provides a measurement of the momentum, \ptot, of charged particles with
a relative uncertainty that varies from 0.5\% at low momentum to 1.0\% at 200\gev.
Natural units with $\hbar=c=1$ 
are used throughout.
The impact parameter (IP), defined as 
the minimum distance of a track to a primary vertex (PV),
is measured with a resolution of $(15+29/\pt)\mum$,
where \pt is the component of the momentum transverse to the beam, in\,\gev.
Muons are identified by a
system composed of alternating layers of iron and multiwire
proportional chambers.

This analysis uses \pPb collision data at a centre-of-mass energy 
per nucleon pair of \sqsnn=8.16\tev recorded in 2016 by the \lhcb detector,
with an energy of $6.5\tev$ for the $p$ beam and $2.56\tev$ per nucleon for the Pb beam.
Since the collision system is asymmetric, 
a detector acceptance of $2.0<\eta<4.5$ in the laboratory frame is 
translated into a muon rapidity ($\murapstar$) acceptance of $1.53<\murapstar<4.03$ 
in the centre-of-mass frame\,\footnote{When the momentum of a particle is much greater than its mass, its rapidity is approximately equal to its pseudorapidity ($y\sim\eta$).},
when the $p$ beam travels towards the positive $\eta$ direction of the \lhcb detector (forward collision).
Similarly, when the $p$ beam travels in the negative $\eta$ direction (backward collision),
the muon rapidity acceptance in the centre-of-mass frame is 
$-4.97<\murapstar<-2.47$.
The integrated luminosity is determined~\cite{LHCb-PAPER-2014-047} to be $12.2\pm0.3\invnb$ for forward collisions
and $18.6\pm0.5\invnb$ for backward collisions. 
For the background and efficiency estimation, 
$pp$ collisions 
recorded in 2016 at \sqs=13\tev are used, corresponding to an integrated luminosity of 1.7\invfb.

Full detector simulation is required to model the detector acceptance effects and 
selection efficiencies. 
The \pPb collisions are simulated by
embedding multiple minimum-bias events generated separately 
into a \decay{\Z}{\mumu} signal event,
so that the multiplicity profile in the simulated
events agrees with that of the collision data. 
The signal \decay{\Z}{\mumu} events are generated with
\pythia 8~\cite{Sjostrand:2007gs,Sjostrand:2006za} using a specific \lhcb configuration and the CTEQ6L1~\cite{Pumplin:2002vw} PDF set
 with a specific \lhcb configuration~\cite{LHCb-PROC-2010-056}
assuming $pp$ interactions with beam momenta equal
to the momenta per nucleon of the $p$ and Pb beams.  
Decays of unstable particles
are described by \evtgen~\cite{Lange:2001uf}, in which final-state
radiation is generated using \photos~\cite{Golonka:2005pn}. 
Minimum-bias events are generated using
the \textsc{Epos} event generator with
the \lhc model~\cite{Pierog:2013ria}.
The interaction and the response of the generated particles with the detector
are implemented using the \geant toolkit~\cite{Allison:2006ve, Agostinelli:2002hh}, 
as described in Ref.~\cite{LHCb-PROC-2011-006}.

The results of the analysis are compared to theory
predictions determined at fixed order in pQCD and given by the \powheg v2~\cite{Nason:2004rx, Frixione:2007vw, Alioli:2010xd, Alioli:2008gx} generator,
which calculates the hard interaction using pQCD at NLO.
The \powheg package is interfaced with the \lhapdf package~\cite{Buckley:2014ana} to
supply different (n)PDF sets for describing the
non-perturbative initial states.
PDF sets CT14~\cite{Dulat:2015mca} and
CTEQ6.1~\cite{Stump:2003yu} are
used for the $p$ side of the \pPb collisions,
and nPDF sets
EPPS16~\cite{Eskola:2016oht,Dulat:2015mca}
and nCTEQ15~\cite{Kovarik:2015cma}
are used for the Pb side.
Choosing CT14 and CTEQ6.1 PDF sets for the $p$ 
side is motivated by the fact that 
they are the free proton PDFs used inside
the EPPS16 and nCTEQ15 nPDF sets, respectively.
The muon rapidity acceptances
in $pp$, in forward and in backward \pPb collisions are different,
which need to be corrected for the \rfb and \rpa measurements.
The \powheg generator with the CTEQ6.1 proton PDF is 
used to derive the corresponding correction factors.

\section{Event selection}
\label{selection}

The online event selection is performed by a trigger,
which consists of a hardware stage followed by a
two-level software stage.
The hardware trigger used in this analysis selects events
containing at least one muon with \pt greater than 500\mev.
In the first stage of the software trigger,
at least one muon with \pt greater than 1.3\gev and
$p$ greater than 6.0\gev is required.
Alignment and calibration
of the detector are performed in near real-time~\cite{LHCb-PROC-2015-011}.
The same alignment and calibration information is propagated to the offline reconstruction,
ensuring consistent and high-quality particle identification information between
the trigger and the offline software.
The present analysis is performed directly 
using candidates reconstructed at the trigger stage~\cite{LHCb-DP-2012-004,LHCb-DP-2016-001}, 
owing to the identical performance of the 
online and offline reconstruction.

\begin{figure}[hbpt]
  \begin{center}
    \includegraphics[width=0.49\linewidth]{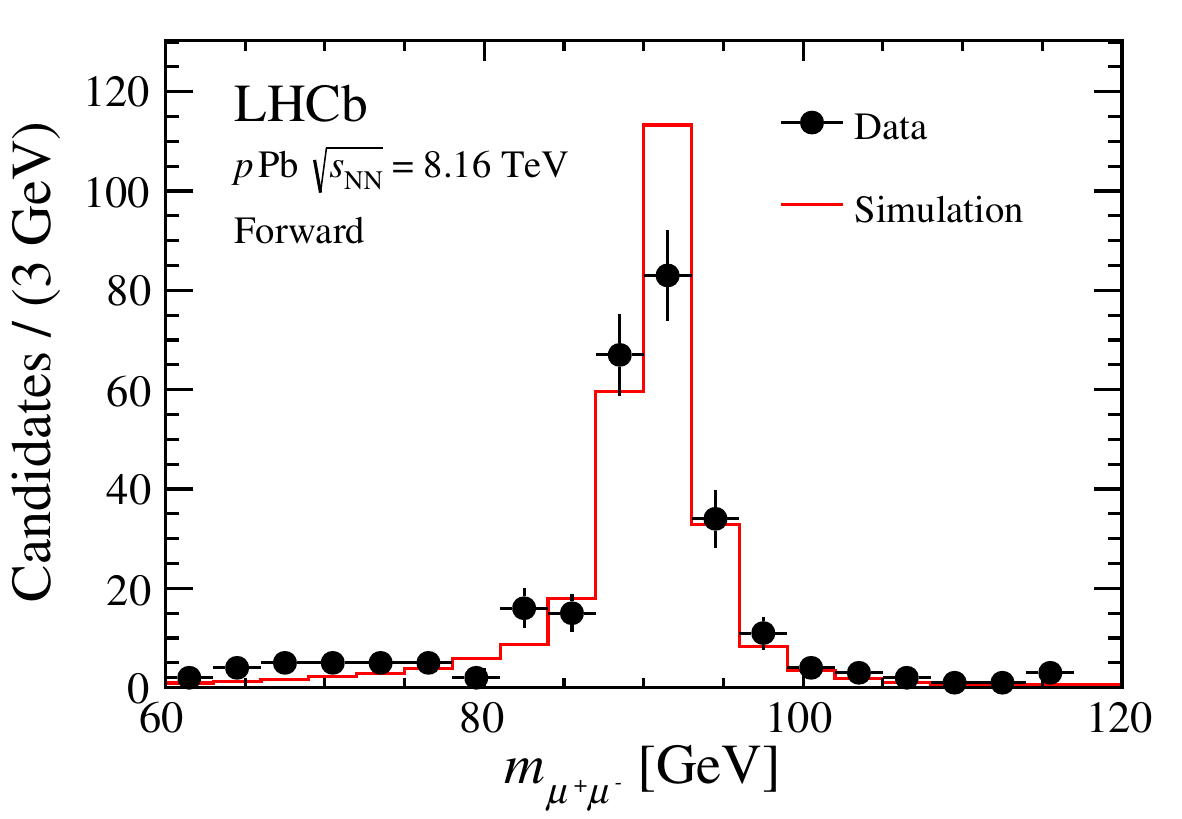}
    \put(-170,70){(a)}
    \includegraphics[width=0.49\linewidth]{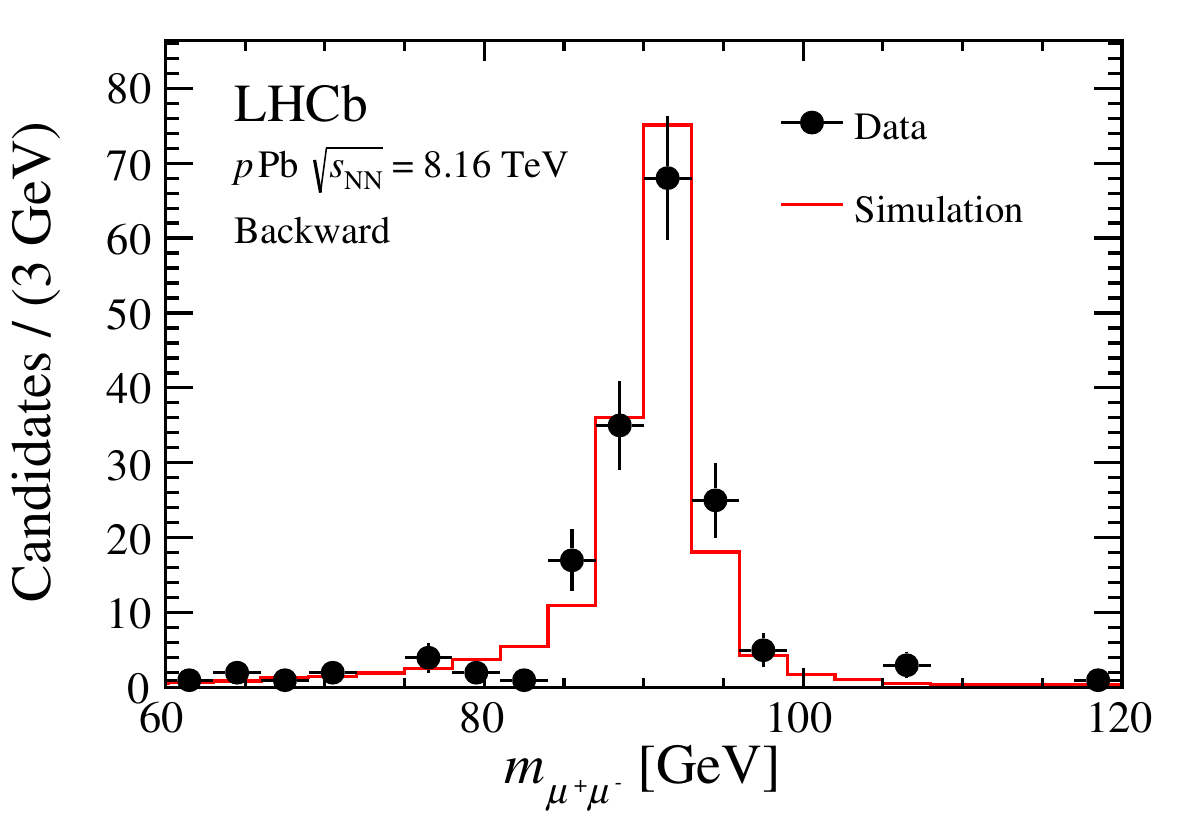}
    \put(-170,70){(b)}
    \vspace*{-0.5cm}
  \end{center}
\caption{Dimuon invariant mass of the selected events for 
(a) forward and (b) backward collisions, respectively.
The red histograms show the distributions from
simulation normalised
to the number of observed candidates.}
  \label{fig:candidates}
\end{figure}

The second software-trigger stage and the offline reconstruction level
selection require each event to have a PV
reconstructed using 
at least four tracks measured in the vertex detector.
For events with multiple PVs, the PV that has the smallest \chisqip
with respect to the selected 
dimuon candidate is chosen, where 
\chisqip is defined as the difference between the vertex-fit \chisqndf calculated
with and without the two tracks from the dimuon candidate included in the vertex fit.
Each identified muon candidate is required to 
have $\pt>20\gev$, $2<\eta<4.5$ and
a good track fit.
The two muon tracks of the \Z boson candidate must form a good-quality vertex (vertex-fit $\chisqndf<25$), 
representing a tighter selection compared to the software trigger requirement.
The dimuon invariant mass is required to be in the range between 60 and 120\gev.
The selected numbers of candidates are 
$268$ and $166$
for forward and backward collisions, respectively.
The dimuon invariant mass of the selected candidates is shown in Fig.~\ref{fig:candidates}
 for both forward and backward collisions.
The red histogram shows the distributions of simulated \Zmumu events generated 
using \pythia\,8 with the
CTEQ6L1~\cite{Pumplin:2002vw} PDF set, normalized
to the number of observed candidates.

\section{Data analysis}

This analysis measures the cross-section of the $\Z\to\mumu$ production
in a fiducial region with $60 < m_{\mumu} < 120\gev$ where both muons satisfy
$\pt>20~\gev$ and $2 < \eta < 4.5$.
The differential cross-section in the fiducial region is defined as
\begin{equation}
\label{eq:xsec_diff}
\frac{\deriv\fidxsec_{\Zmumu}}{\deriv x} = \frac{\rho(x)\cdot \fsrcorr}{\lum\cdot\effreco(x)\cdot\effmuid(x)\cdot\efftrig(x)}
 \cdot \frac{\deriv \ncand}{\deriv x}\,,
\end{equation}
where \ncand is the number of observed candidates after the selection in the fiducial region;
$\rho$ is the purity (the fraction of signal events);
\fsrcorr is the correction for final state radiation (FSR);
\lum is the integrated luminosity;
 \effreco, \effmuid and \efftrig are the efficiencies of
reconstruction and selection, muon identification and 
trigger selection, respectively;
and $x$ 
can be \zrapstar,
\ZpT or \phistar 
(applied throughout this section unless stated otherwise).
The variable \phistar is defined as
\begin{equation}\label{phi}
\phistar = \frac{\tan(\phi_{\mathrm{acop}}/2)}{\cosh(\Delta\eta/2)},
\end{equation}
where $\Delta \eta$ is the difference between the pseudorapidity of the two muons, $\phi_{\mathrm{acop}} = \pi - |\Delta\phi|$ is the acoplanarity, 
with $\Delta \phi$
being the difference between the azimuthal angles of the two muons.
The angle \phistar was introduced and studied in Refs.~\cite{Vesterinen:2008hx,Banfi:2010cf} 
as an observable complementary to the variable \ZpT to study the shape of the \Z boson 
transverse-momentum distribution
with reduced sensitivity to experimental resolution effects.

The ratio of the \Zmumu production  cross-sections between forward
and backward directions, \rfb, is particularly
sensitive to nuclear effects.
In the case of $pp$ collisions,
the \rfb is expected to be unity, because
the \Z boson production cross-section should be symmetric
between forward and backward rapidity.
However, in the case of \pPb collisions,
the nuclear modifications are different
for forward (small Bjorken-$x$, \ie,  $x<10^{-3}$)
and backward (large Bjorken-$x$, \ie, $x>10^{-1}$) rapidity.
As such, there is suppression
at small Bjorken-$x$
due to nuclear shadowing effects~\cite{Brodsky:1989qz},
and enhancement
at large Bjorken-$x$
because of the EMC effect~\cite{EuropeanMuon:1983wih}.
Therefore,
the \rfb is an effective observable to probe
nuclear-matter effects.

The forward-backward ratio is determined 
in the common rapidity range $2.5< |\zrapstar|< 4.0$ as
\begin{equation}
\label{eq:rfb}
\rfb(x)=\left.\frac{\deriv\sigma_{\mathrm{fw}}/\deriv x}{\deriv\sigma_{\mathrm{bw}}/\deriv x}\cdot \kfb(x) \right|_{2.5<|\zrapstar|<4.0}~,
\end{equation}
where $\sigma_{\mathrm{fw}}$ is the cross-section
in the forward collisions, and
$\sigma_{\mathrm{bw}}$ in the backward collisions.
The factor $\kfb(x)$, which corrects for the difference of
the muon rapidity acceptance between forward and backward collisions,
is determined as
\begin{equation}
\kfb(x) = \dfrac{\deriv\sigma'_{(pp,-4.97<\murapstar<-2.47)}/\deriv x}{\deriv\sigma'_{(pp,1.53<\murapstar<4.03)}/\deriv x}\Bigg|_{2.5<|\zrapstar|<4.0}~,
\end{equation}
using \powheg with the proton PDF CTEQ6.1.
Choosing the proton PDF instead of the nPDF avoids the acceptance correction factor being biased by nuclear modification information encoded in the nPDF.
The prime symbol (``$\sigma'$'') in the equation 
indicates that this cross-section is calculated theoretically instead of from data measurements.

The nuclear modification factor (\rpa) 
for forward and backward collisions
is measured as
\begin{equation}
\rpa^{\rm fw}(x)= \kpa^{\rm fw}(x) \cdot \dfrac{\deriv\sigma_{(\pPb, 1.53<\murapstar<4.03)}/\deriv x}{ 208 \cdot \deriv\sigma_{(pp,2.0<\murapstar<4.5)}/\deriv x}
\label{eq:rpa_fw_bins}
\end{equation}
and
\begin{equation}
\rpa^{\rm bw}(x)= \kpa^{\rm bw}(x) \cdot \dfrac{\deriv \sigma_{(\pPb, -4.97<\murapstar<-2.47)}/\deriv x}{ 208 \cdot \deriv \sigma_{(pp,-4.5<\murapstar<-2.0)}/\deriv x}~,
\label{eq:rpa_bw_bins}
\end{equation}
respectively, 
where 208 is the number of binary 
nucleon-nucleon collisions in \pPb collisions. 
The \kpa factor is to correct for the different
\murapstar acceptance between \pPb and $pp$ collisions
and can be similarly calculated using \powheg with the proton PDF set CTEQ6.1, 
\begin{equation}
\kpa^{\rm fw}(x)= \dfrac{\deriv\sigma'_{(pp,2.0<\murapstar<4.5)}/\deriv x}{\deriv\sigma'_{(pp,1.53<\murapstar<4.03)}/\deriv x}
\end{equation}
and
\begin{equation}
\kpa^{\rm bw}(x)= \dfrac{\deriv\sigma'_{(pp,-4.5<\murapstar<-2.0)}/\deriv x}{\deriv\sigma'_{(pp,-4.97<\murapstar<-2.47)}/\deriv x}
\end{equation}
for forward and backward collisions, respectively.
By construction, the nuclear modification factor is expected
to be unity if no nuclear matter effects are present, and can be used to extract nuclear modifications when a proton or neutron is confined in
the Pb nucleus.

The dominant background 
contribution in this analysis is from QCD processes. 
This background has contributions from
heavy flavour hadrons decaying to muons
and charged hadrons mis-identified as muons. 
Contributions from other sources, 
such as \ttbar, \Wp\Wm and \decay{\Z}{\tautau} processes, 
are found to be negligible~\cite{LHCb-PAPER-2015-001}, 
and are not considered in the present work.
Two methods are employed to estimate the QCD background, using $pp$ collision samples
weighted to forward or backward \pPb collisions 
according to the multiplicity profiles.

The first method is a ``same-sign'' technique,
based on the property that the yields of positively 
and negatively charged hadrons are equal in size in the final states.
Therefore, the amounts of same-sign
and opposite-sign muon-pair candidates that originated from charged hadrons
(either mis-identified as muons, or decayed into muons) 
are the same. 
By selecting same-sign muon-pair candidates, the amount of QCD background
contaminating the selected signal
can be determined.

The second method is the ABCD-likelihood technique,
see, \eg,~Refs.~\cite{CDF:1990kbl,ATLAS:2015itk,Kasieczka:2020pil}, 
which is widely used
for QCD background estimation in 
beyond-Standard-Model searches. 
In this approach 
two discriminating variables,
the vertex-fit \chisqndf and muon isolation variable
\Iso, are considered.
The vertex-fit \chisqndf is used to verify if the
pair of muon candidates originates from the same \Z decay,
otherwise it is classified as QCD background.
The muon isolation variable is defined as the fraction of \pt
carried by the muon candidate over the sum of \pt carried by all the
tracks in a cone in the $\eta-\phi$ plane with a size of $\Delta R= \sqrt{{\Delta \eta}^2+{\Delta \phi}^2}<0.5$ around the muon candidate.
This variable is used to assess if a muon candidate is surrounded by charged tracks,
which characterises a mis-identified hadron or a muon from a charged-hadron decay. A requirement of $\Iso>0.7$ can effectively separate the signal from the QCD background~\cite{LHCb-PAPER-2014-022}.
These two variables are used to separate the dimuon sample into four regions:
\begin{itemize}
\item[(A)] vertex-fit $\chisqndf<25$ and at least one muon with $\Iso>0.7$,
\item[(B)] vertex-fit $\chisqndf>70$ and at least one muon with $\Iso>0.7$,
\item[(C)] vertex-fit $\chisqndf<25$ and both muons with $\Iso<0.7$, and
\item[(D)] vertex-fit $\chisqndf>70$ and both muons with $\Iso<0.7$,
\end{itemize}
where ($\mathrm{A+C}$) is the signal selection fiducial volume.
The background contamination in the region ($\mathrm{A+C}$) is therefore
estimated as
$N_\mathrm{bkg}(\mathrm{A+C}) = N(\mathrm{C})/N(\mathrm{D})\cdot N(\mathrm{B+D})$.

While the ABCD method is more sensitive 
to QCD backgrounds from heavy-flavour decays, 
and the same-sign method to hadron
mis-identification, 
there is a sizable overlap 
between the two methods. 
Consequently, the final QCD background is taken 
as the sum of the results from the two methods
with a subtraction of this duplication.
An estimation of the duplication of 
the two methods is
performed by using the same-sign sample as input 
to the ABCD method.
The resulting average purity is
$(99.70\pm0.07)\%$ and $(99.75\pm0.08)\%$ for the
forward and backward collisions, respectively.
For the differential measurements as functions of
\zrapstar, \ZpT, and \phistar, 
the background is estimated in the same way and found to be 
independent of these variables.
The average purities given above 
for forward and backward collisions
are used correspondingly for the differential measurements.

The efficiency cannot be directly derived from the 
\pPb dataset because of its small statistical size. 
The high statistical $pp$ collision dataset is
instead chosen as the starting point for efficiency studies.
The efficiencies are evaluated in the fiducial 
region in two steps. 
In the first step, 
a tag-and-probe method~\cite{LHCb-DP-2013-002} is used to derive the efficiencies
as a function of 
multiplicity, \mupt, \meta,
\ZpT and \Z boson rapidity in the lab frame (\zrap) 
from both $pp$ collision data
and \Zmumu simulation.
The ratios between data and simulation
are taken as correction functions.
In the second step, 
the \pPb simulation is weighted to the
\pPb collider data according to the corresponding
multiplicity profiles.
The correction functions derived in the first step 
are then used to further 
weight the \pPb simulation 
to correct for the differences between
data and simulation. 
After a series of 
reweightings, 
the \pPb simulation
can correctly reproduce the efficiencies of the
\pPb data.
The final efficiency to be used for cross-section measurements
can be then derived as the fraction of the weighted yields passing the corresponding selection criteria using the weighted \pPb simulation.

\begin{table}[b]
\begin{center}
\caption{Observed number of candidates, input values and uncertainties for the overall cross-section and the theory corrections with uncertainties for the \rfb and \rpa measurements.}
\label{tab:system-uncertainties-sources}

\begin{tabular}{p{4cm}ll}
\hline\hline
Quantity           & Forward             & Backward \\
\hline
$N_{\text{cand}}$ (for \fidxsec)   &$268\pm16$  & $166\pm13$ \\
$N_{\text{cand}}$ (for \rfb) &$160\pm13$        &$166\pm13$  \\
$N_{\text{cand}}$ (for \rpa) &$241\pm16$       &$166\pm13$  \\ \hline
$\rho\,[\%]$    &$99.69\pm0.07$       &$99.75\pm0.08$ \\
\effreco\,[\%]      &$87.2\pz\pm2.9$      &$72.0\pz\pm2.5$ \\
\effmuid\,[\%]      &$97.3\pz\pm0.3$      &$97.3\pz\pm0.3$ \\
\efftrig\,[\%]      &$98.3\pz\pm0.6$      &$97.1\pz\pm0.6$ \\ 
\lum\,$[\invnb]$      &$12.2\pz\pm0.3$      &$18.6\pz\pm0.5$ \\
\hline
\fsrcorr            &$1.025\pm0.001$     &$1.025\pm0.001$ \\
\kfb (for \rfb)    &$0.65\pz\pm 0.02$ & \pz\pz--\\
\kpa (for \rpa)    &$0.706\pm 0.002$ & $1.518\pm 0.003$ \\
\hline\hline
\end{tabular}
\end{center}
\end{table}

The resulting average efficiencies (\effreco, \effmuid, and \efftrig) are shown in Table~\ref{tab:system-uncertainties-sources} for the 
forward and backward collisions. 
The much smaller \effreco efficiency in the 
backward collisions is mainly due to the far larger 
contamination from charged tracks when the debris of the Pb nucleus 
travels towards the \lhcb detector.
The \effreco efficiency for the differential measurements varies
from $66.6\%$ to $91.2\%$, depending on the \zrapstar, \ZpT or \phistar variables.
The efficiencies \effmuid and \efftrig show negligible dependencies on these variables
and are considered as constants.

The detector resolution can lead to migration of
events from one interval to another in
the distribution of an observable.
The migration effect is studied 
by examining the ratio of the 
distributions 
of the \zrapstar, \ZpT and \phistar observables
between generated and reconstructed events
using simulation.
It is found to be negligible 
for the observables \zrapstar and \phistar,
thanks to the excellent angular resolution
of the \lhcb detector. 
However, for the variable \ZpT 
the migration effect
can be as large as $\sim 10\%$. 
An unfolding correction is applied 
for the \ZpT distributions
using the RooUnfold~\cite{Adye:2011gm} software package
with an iterative Bayesian approach~\cite{DAgostini:1994fjx}.

Final-state radiation effects can shift certain events out of the kinematic
acceptance boundaries. 
For a fair comparison of the results with 
theoretical predictions,
the FSR correction (\fsrcorr) is applied to the measured result, 
which is defined as the ratio of the cross-sections
with and without the FSR effect turned on
calculated using \powheg together with \pythia8~\cite{Sjostrand:2007gs}.
The average value of the FSR correction 
is about 2.5\%.

For the nuclear modification factor measurement,
the $pp$ reference cross-section at $8.16\tev$ 
is interpolated using a third-order polynomial function 
from the 
$pp$ measurements at 7, 8, and
13\tev~\cite{LHCb-PAPER-2015-001, LHCb-PAPER-2015-049, LHCb-PAPER-2016-021, LHCb:2021huf} by the \lhcb experiment.
The resulting interpolated total cross-section at
8.16\tev is $(98.1\pm0.6)\pb$ in the rapidity range
$2.0 < \zrapstar <4.5$, 
and varies little with respect to the measured central value of 95\pb at 8\tev~\cite{LHCb-PAPER-2015-049}.
The differential cross-section in intervals of \zrap, \ZpT and \phistar variables
is also interpolated in the same way for each interval.
The method has been cross-checked via a linear extrapolation 
using the 7 and 8\tev points 
leading to a negligible difference ($<0.1\%$) in the result.

\section{Systematic uncertainties}
\label{sec:systematic}

Sources of systematic uncertainties 
considered are the 
background estimation, efficiency modelling, 
FSR correction, detector resolution correction,
integrated luminosity, and theoretical corrections.

The uncertainty from the background 
estimation is less than $0.1\%$, and 
includes
the statistical uncertainty of the $pp$ collision dataset used in the estimation
and the multiplicity weighting.
The multiplicity weighting uncertainty is estimated
by varying the weighting function within an envelope defined by 
the statistical fluctuations of the multiplicity profiles of the $pp$ and 
\pPb datasets.
 The efficiency uncertainty is about $2.5\%$ on average and
 consists of statistical uncertainties from the
 $pp$ collision dataset and the simulation samples of forward and backward collisions used in the tag-and-probe method,
 together with the multiplicity weighting uncertainty. 
 Here the multiplicity weighting uncertainty is estimated in the 
 same way as for the background.
 The FSR correction uncertainty is estimated to be less than $0.1\%$ 
as the difference of the default correction factors derived using \powheg with \pythia8 with respect to an alternative approach using \photos~\cite{davidson2015photos}.
 The uncertainty of the $pp$ reference cross-section at 8.16\tev is interpolated to be 0.6\% on average as an error band enclosed by the upper and lower edges of the error bars of the 7, 8 and 13\tev measurements.
The luminosity uncertainty is about $2.5\%$~\cite{LHCb-PAPER-2014-047}.
For the differential measurements of the cross-section, \rfb, and \rpa as 
a function of the \ZpT, the detector-resolution correction is applied. 
The uncertainty is given by the difference between the Bayesian
approach and other unfolding techniques~\cite{Hocker:1995kb, cowan1998statistical},
which are considered only in the differential measurements. 

For the \rfb and \rpa measurements,
uncertainties from the \murapstar acceptance corrections 
\kfb and \kpa are considered, 
which include PDF uncertainties, and factorisation and renormalisation scale uncertainties.

The uncertainties of the theoretical predictions shown
in the comparisons include the PDF and nPDF uncertainties, and the factorisation and renormalisation scale uncertainties. 
The (n)PDF uncertainties are estimated by varying the eigenvectors of
the (n)PDF sets up and down simultaneously on both the $p$ side and the Pb side. 
The factorisation and renormalisation scale uncertainties are obtained by varying the scale
factors from 0.5 to 2.0. 
All the theoretical uncertainties are converted to a $68\%$ confidence level when comparing
with the measurements.

Table~\ref{tab:system-uncertainties-sources}
summarises the inputs and their uncertainties
for the total cross-section,
\rfb and \rpa measurements.
The uncertainties are evaluated separately
for \rfb and \rpa measurements because of the different
\zrapstar coverage between $pp$, forward and backward \pPb collisions, and are found to be identical to that 
for the cross-section measurement.

For the differential measurements of the cross-section, \rfb, and \rpa as
a function of the \zrapstar, \ZpT and \phistar variables, 
the uncertainty from the integrated luminosity is considered
 to be $100\%$ correlated, while the uncertainties from other sources are considered to be partially correlated.
Correlation matrices are evaluated for the statistical uncertainty
and for the reconstruction and selection efficiency uncertainty,
since they form the two largest sources of uncertainty.
The statistical uncertainty correlation due to interval migration
is evaluated using simulation
by comparing the distributions at generator level and reconstruction level.
As a result, a $10$\,--\,$26\%$ correlation can be observed for \ZpT,
a $2$\,--\,$3\%$ correlation for \zrapstar, and a negligibly small correlation for \phistar.
The correlation matrices for the reconstruction and selection efficiencies
are calculated by varying the efficiencies in each interval independently within their uncertainties.
Large correlations between different intervals are observed.
The resulting correlation matrices for statistical uncertainty and the
reconstruction and selection efficiency uncertainties are given
in Appendix~\ref{apdx:corr_matrix}.

\section{Results and discussion}
\label{sec:Results}

The total fiducial cross-section of \Zmumu production in forward and backward collisions
is measured to be 
\begin{align*}
\fidxsec_{\Zmumu,~\pPb} &= ~26.9 \pm 1.6 \pm 0.9 \pm 0.7 ~\nb~,\\
\fidxsec_{\Zmumu,~\Pbp} &= ~13.4 \pm 1.0 \pm 0.5 \pm 0.3 ~\nb~,
\end{align*}
where the first uncertainty is statistical,
the second is systematic
and the third is due to the uncertainty in the luminosity determination.
As a result of the different muon rapidity acceptances 
between forward and backward collisions, 
the \zrapstar coverage is also different.
An additional acceptance requirement of
$1.5<\zrapstar<4.0$ for forward collisions
and $-4.0<\zrapstar<-2.5$ for backward collisions
is applied on top of the fiducial volume.

\begin{figure}[htbp]
\begin{center}
\includegraphics[width=0.78\linewidth]{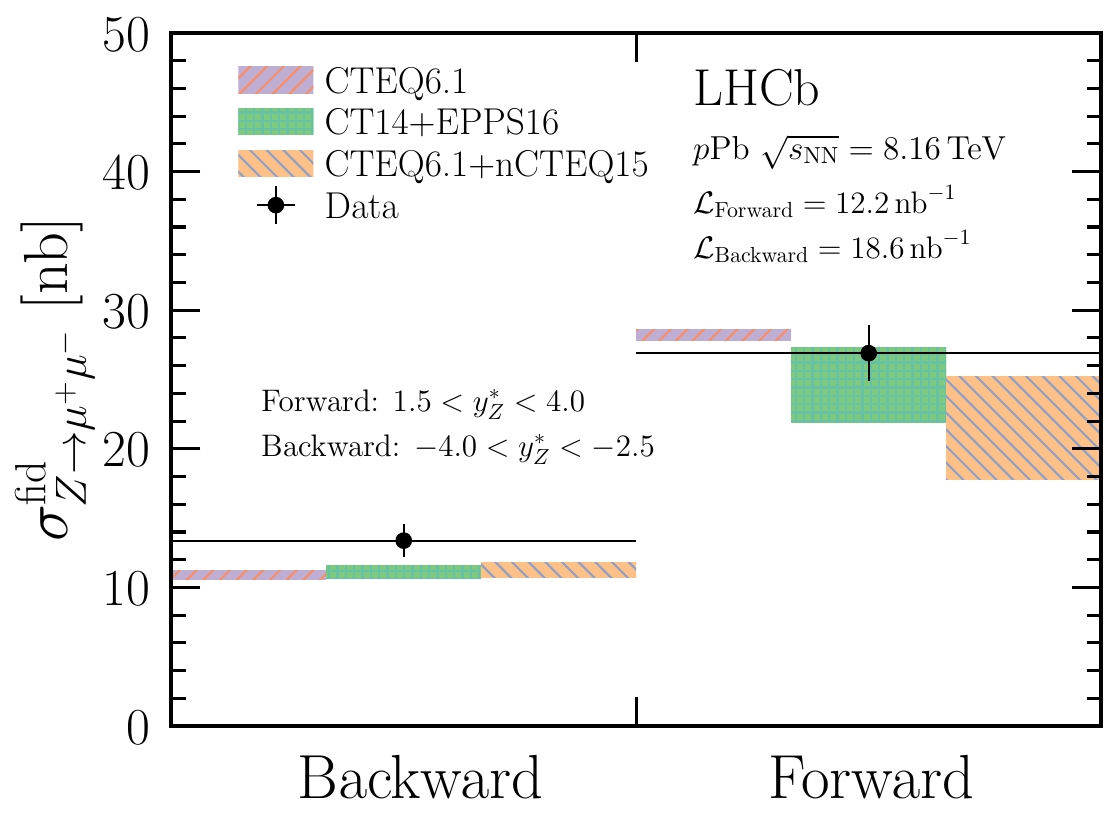}
\vspace*{-0.5cm}
\end{center}
\caption{
The measured overall \Zmumu production fiducial cross-section
compared to the \powheg prediction using
CTEQ6.1, EPPS16 and nCTEQ15 (n)PDF sets,
for forward and backward collisions, respectively.
}
\label{fig:fid-cross-section-result}
\end{figure}

Figure~\ref{fig:fid-cross-section-result} shows the
measured results compared
to the \powheg calculations using the
CTEQ6.1 proton PDF set, EPPS16 nPDF and nCTEQ15 nPDF sets.
For forward collisions, the measured value agrees with theoretical predictions
with a much smaller uncertainty, which 
indicates a strong constraining power to
the theoretical modelling of the nPDFs.
For backward collisions,
the measured value is slightly higher
but statistically compatible with
the theory predictions (difference below 2$\sigma$). 
This trend was also observed in the previous
result at $\sqsnn=5.02\tev$~\cite{LHCb-PAPER-2014-022}.

\begin{figure}[tbp]
\begin{center}
\includegraphics[width=0.48\linewidth]{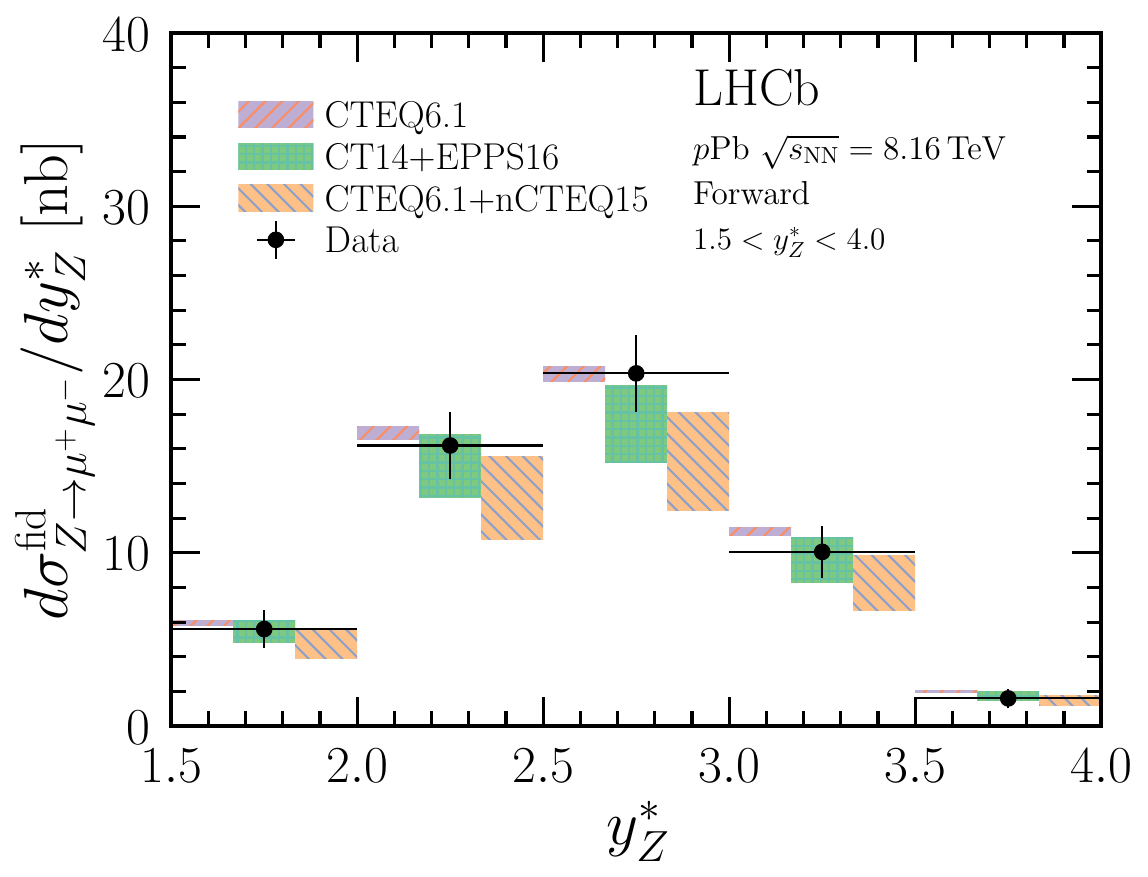}
\put(-50,80){(a)}
\hspace*{0.5cm}
\includegraphics[width=0.48\linewidth]{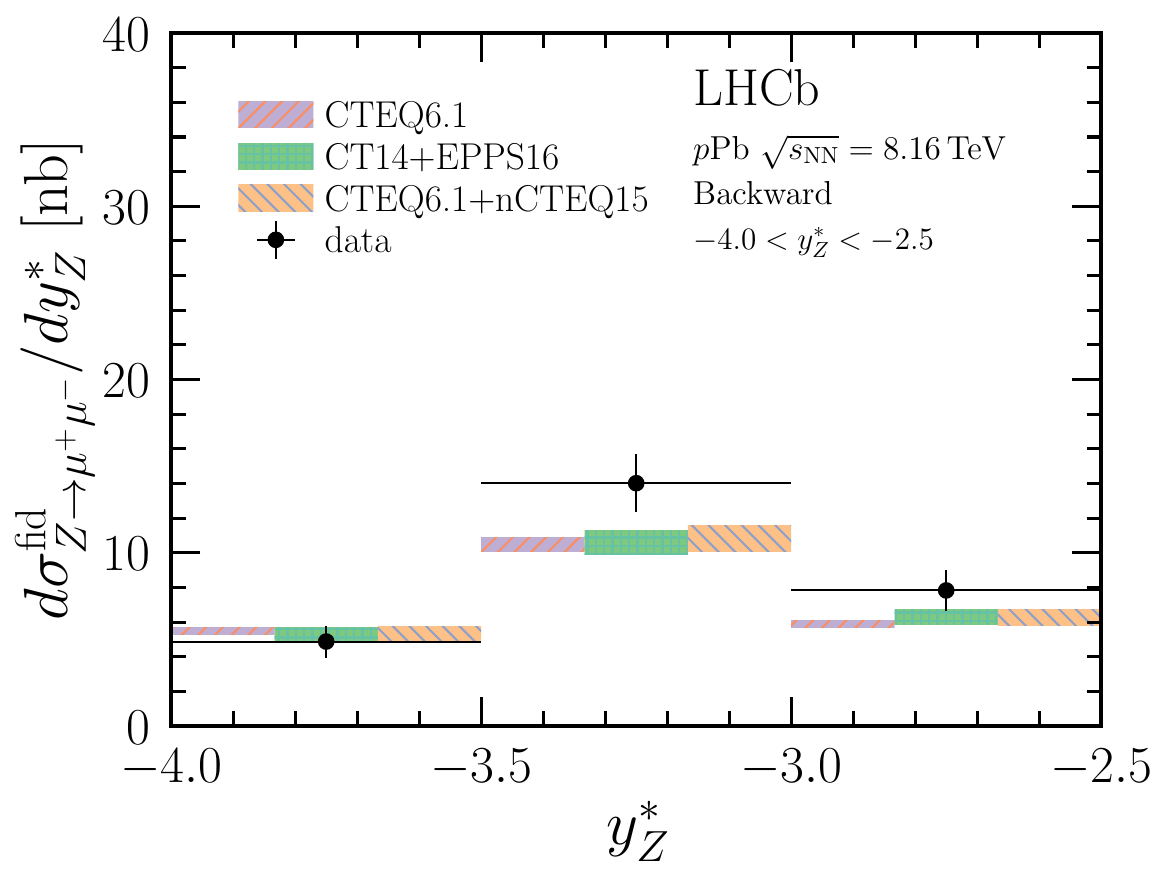}
\put(-50,80){(b)}\\
\end{center}
\caption{
The measured differential fiducial cross-section as a function of \zrapstar for
(a) forward and (b) backward collisions.
The theoretical predictions are calculated using \powheg
with CTEQ6.1, EPPS16 and nCTEQ15 (n)PDF sets.
}
\label{fig:cross-section-zrap}
\end{figure}

The measured differential fiducial cross-sections as a function
of \zrapstar are shown in Figure~\ref{fig:cross-section-zrap} for
forward and backward collisions,
together with the \powheg calculations with
CTEQ6.1, EPPS16 and nCTEQ15 (n)PDF sets.
For forward collisions,
the measured values show good agreement
with the \powheg calculations, with a smaller uncertainty
for the two intervals of $2.0<\zrapstar<3.0$ compared
to the theoretical calculations,
which can be used to further constrain the
nPDFs.
For backward collisions,
the uncertainty of the measurement is larger than
that of the \powheg calculation, and the measured
central value is higher than the prediction especially
for the $-3.5 <\zrapstar <-3.0$ interval by about 2$\sigma$.
However, the measurement and calculation are compatible within uncertainties.

\begin{figure}[tbp]
\begin{center}
\includegraphics[width=0.48\linewidth]{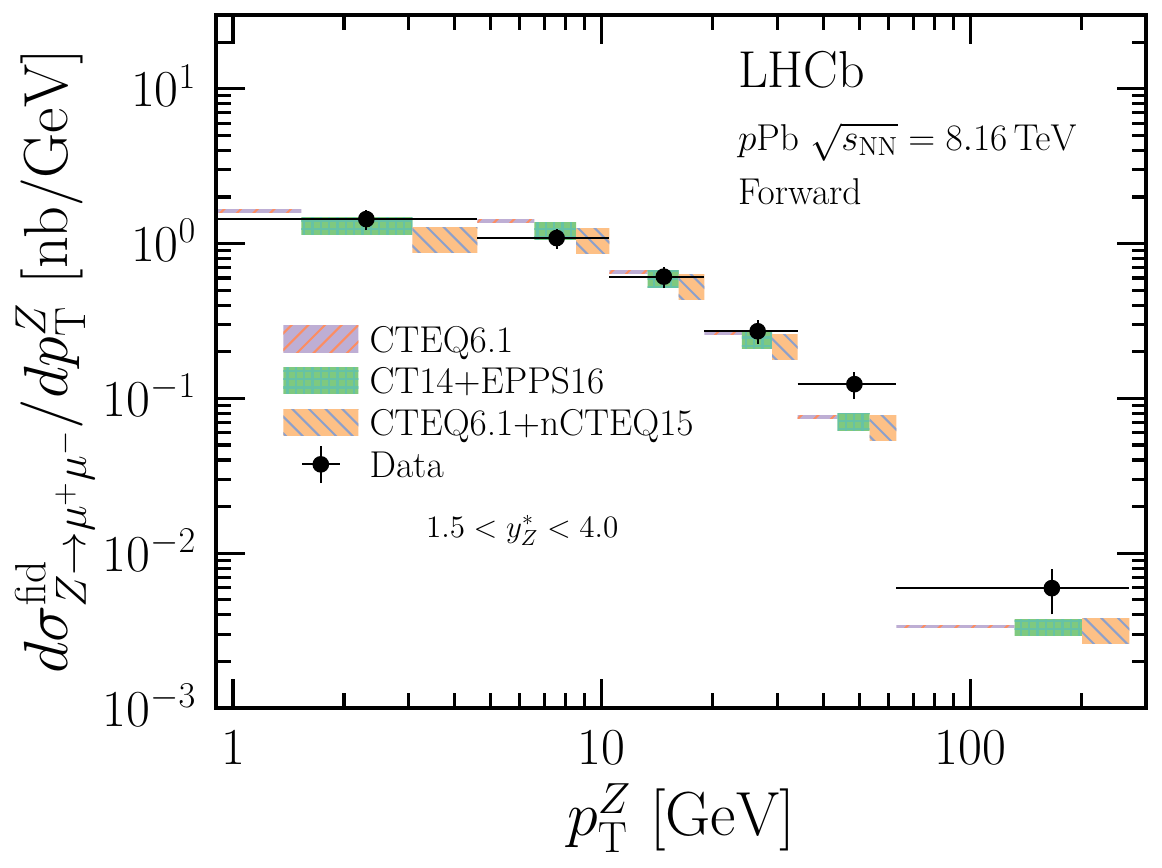}
\put(-40,80){(a)}
\hspace*{0.5cm}
\includegraphics[width=0.48\linewidth]{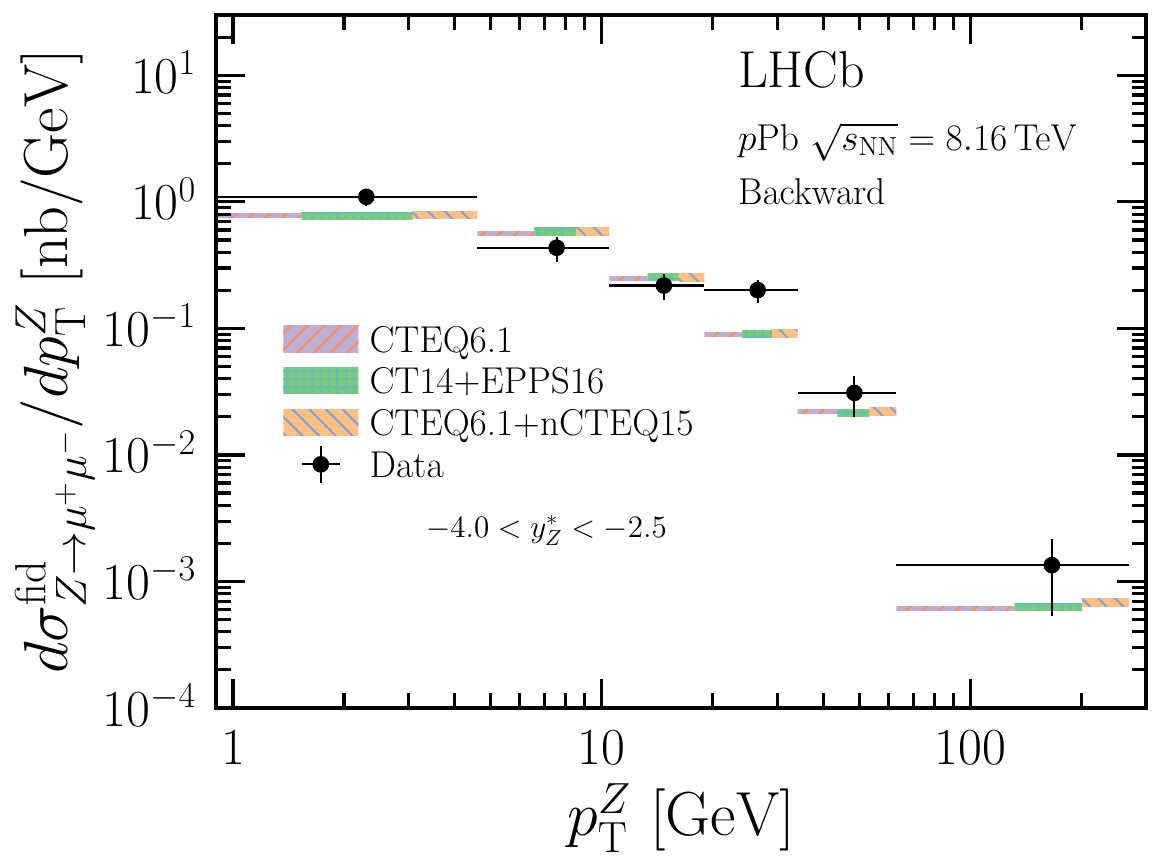}
\put(-40,80){(b)}\\
\vspace*{0.3cm}
\includegraphics[width=0.48\linewidth]{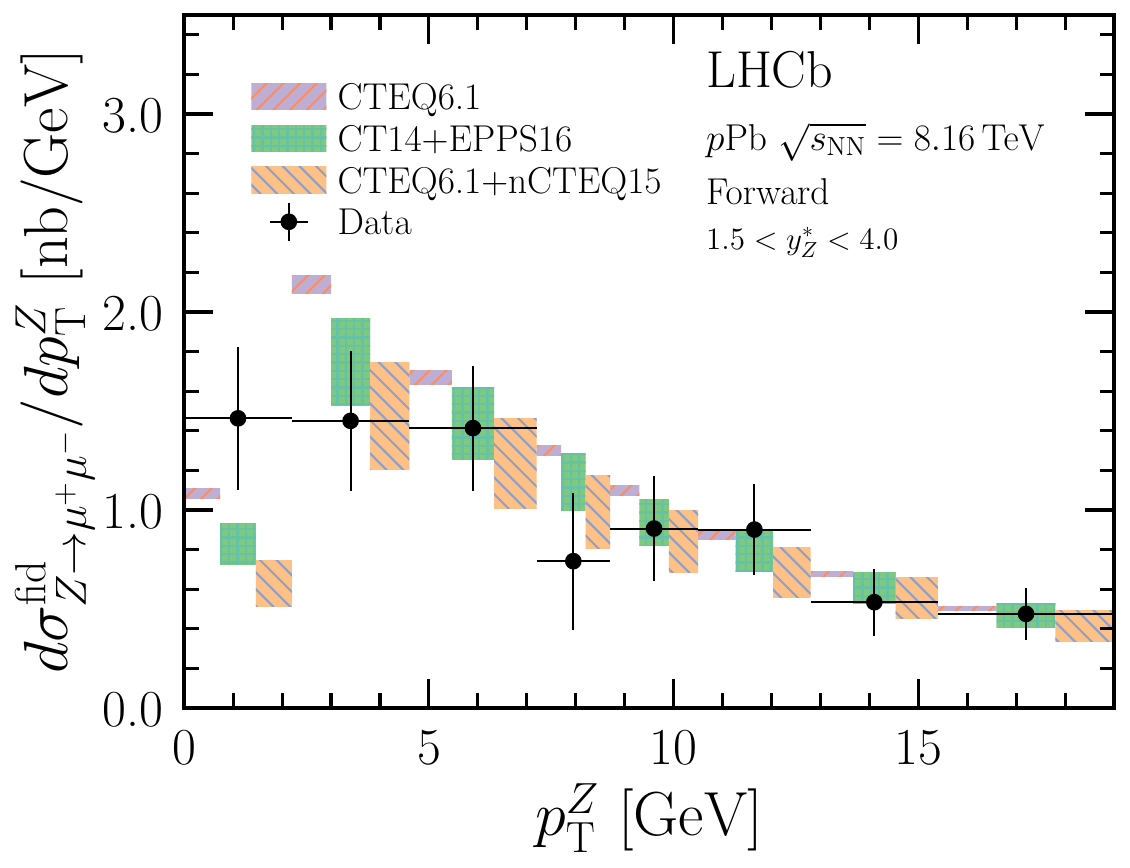}
\put(-40,80){(c)}
\hspace*{0.5cm}
\includegraphics[width=0.48\linewidth]{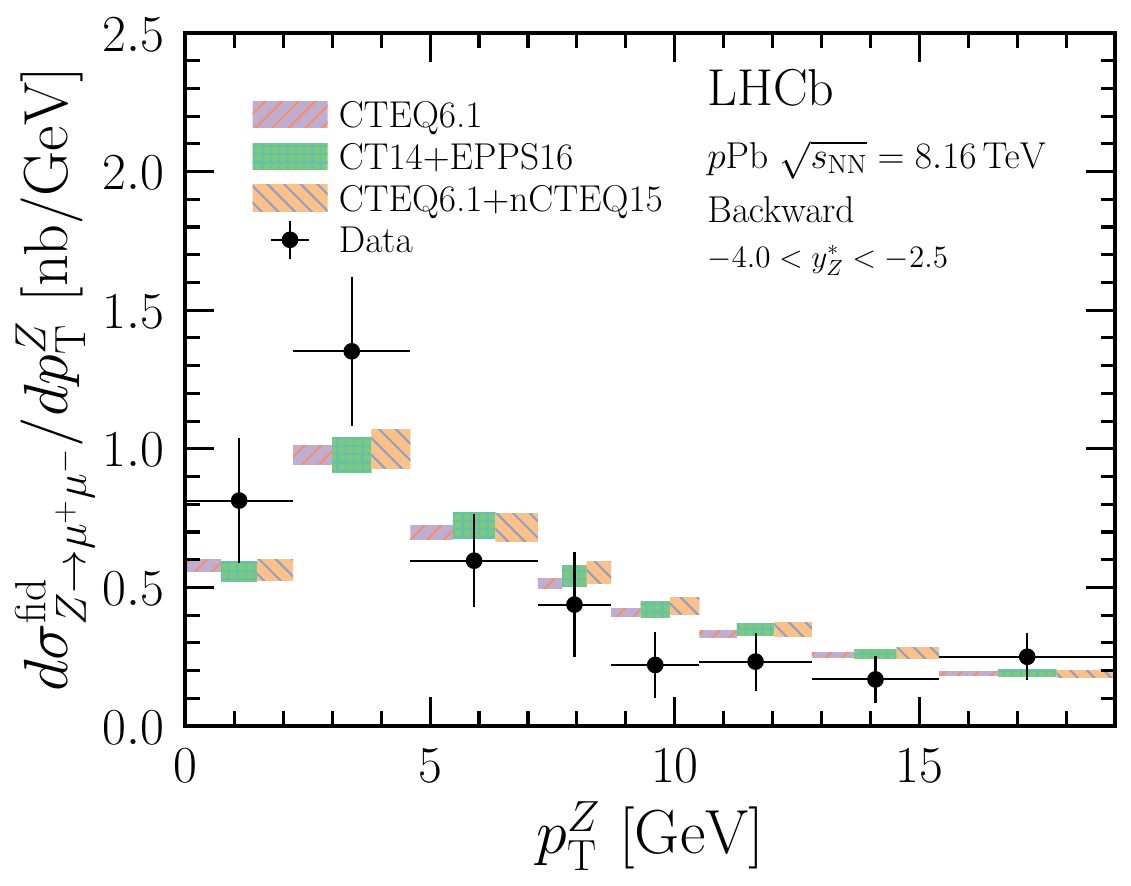}
\put(-40,80){(d)}
\end{center}
\caption{
The measured differential fiducial cross-section as a function
of \ZpT for
(a) forward and (b) backward collisions, and the 
corresponding version for 
(c) forward and (d) backward collisions with fine intervals at low \ZpT.
The theoretical predictions are calculated using \powheg
with CTEQ6.1, EPPS16 and nCTEQ15 (n)PDF sets.
}
\label{fig:cross-section-zpt}
\end{figure}

\begin{figure}[tbp]
\begin{center}
\vspace*{0.3cm}
\includegraphics[width=0.48\linewidth]{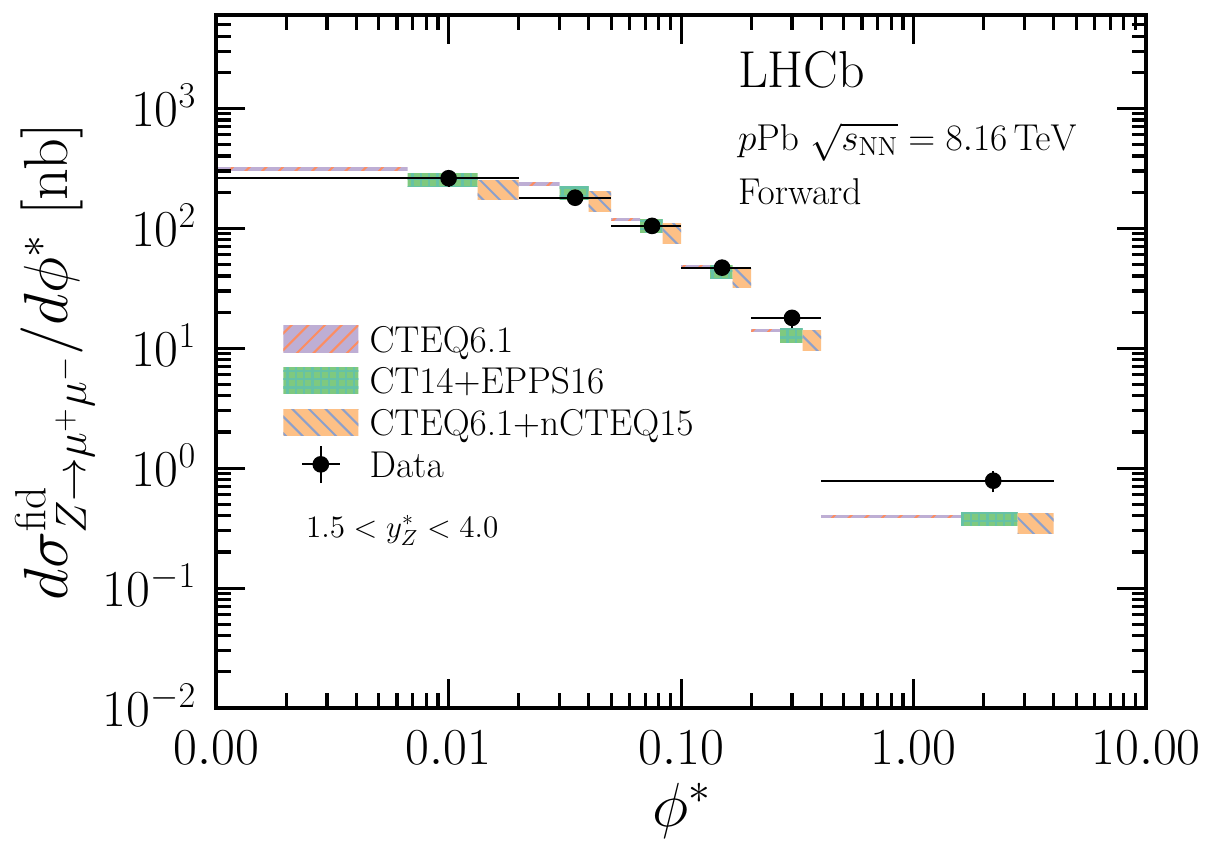}
\put(-50,80){(a)}
\hspace*{0.5cm}
\includegraphics[width=0.48\linewidth]{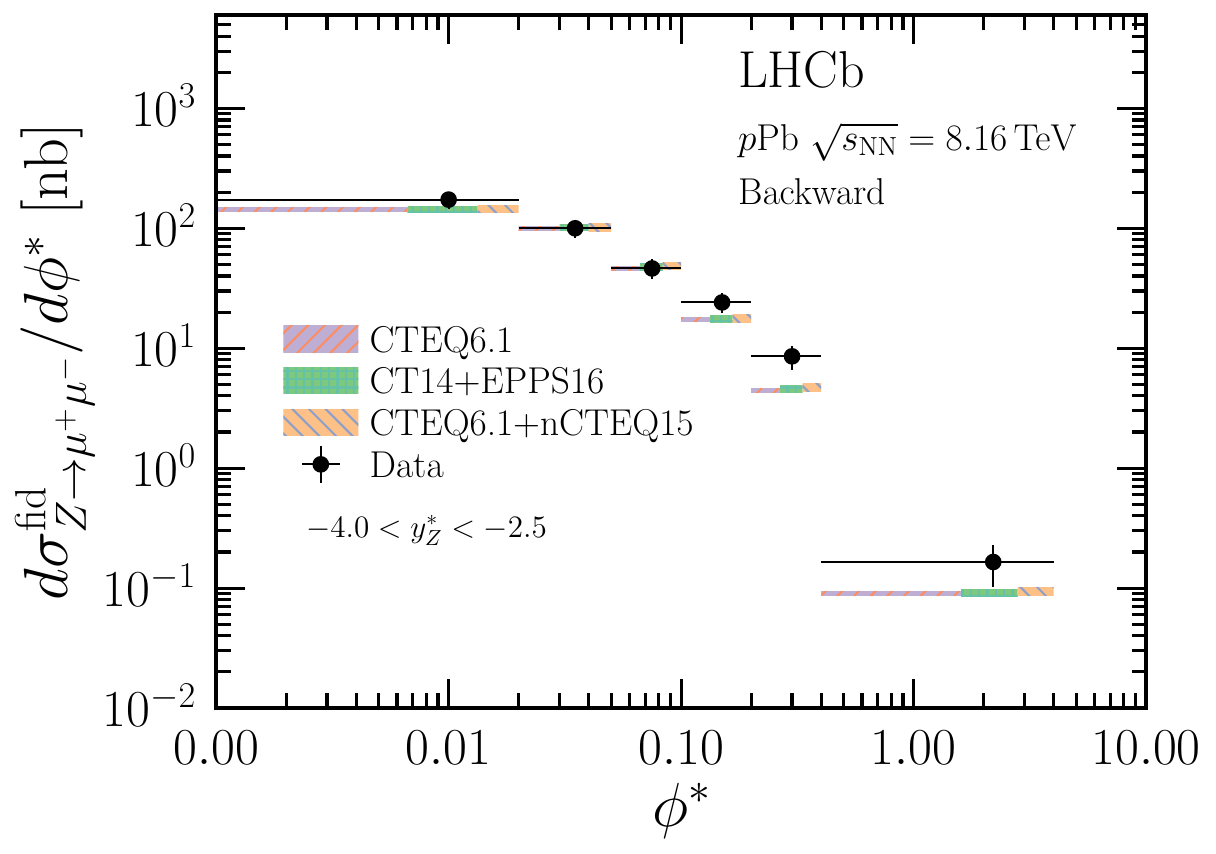}
\put(-50,80){(b)}
\end{center}
\caption{
The measured differential fiducial cross-section as a function of \phistar for
(a) forward and (b) backward collisions.
The theoretical predictions are calculated using \powheg
with CTEQ6.1, EPPS16 and nCTEQ15 (n)PDF sets.
}
\label{fig:cross-section-phistar}
\end{figure}

The differential fiducial cross-section as a function
of \ZpT is shown in
Figure~\ref{fig:cross-section-zpt} (a) and (b),
together with the \powheg calculations.
For forward collisions,
the measured values give a smaller uncertainty compared
to the \powheg calculation for low \ZpT intervals,
showing a strong constraining power.
For backward collisions,
the uncertainty of the measurement is larger than
that of the \powheg calculations
but statistically compatible.
For \ZpT above $30\gev$ the central value of the measurement is
slightly larger than the \powheg calculations, but statistically compatible with them
for both forward and backward collisions.
The same cross-section measurement as a function of
\ZpT at low transverse momentum, where non-perturbation effects
start to dominate, is also provided
with finer intervals,
in order to aid theoretical TMD studies, 
as shown in Figure~\ref{fig:cross-section-zpt} (c) and (d).
The differential
fiducial cross-section as a function
of the \phistar variable is shown in
Figure~\ref{fig:cross-section-phistar}
together with the \powheg calculations.
Since the angular observable \phistar is equivalent to \ZpT but not impacted by the momentum resolution, a similar conclusion can be drawn as for \ZpT.
The detailed numerical values  and uncertainties of the differential cross-section as a function of
\zrapstar, \ZpT and \phistar, together with the corresponding \fsrcorr factors are given in 
Appendix~\ref{apdx:cross-section}.

The cross-sections used to determine the forward-backward ratio are different
with respect to the fiducial cross-sections.
They are measured in the common
\Z-boson rapidity range
$2.5<|\zrapstar|<4.0$ with
efficiency, purity and FSR corrections also determined in the same rapidity
range as shown in Table~\ref{tab:system-uncertainties-sources},
These corrections are found to be identical to those of
the fiducial cross-section measurement.
The resulting cross-sections for \rfb measurements are
$16.1 \pm 1.5\nb$
and
$13.4 \pm 1.2\nb$
for forward and backward collisions, respectively.
To eliminate the impact from different \murapstar acceptances,
the correction factor \kfb for the overall \rfb measurement is calculated to be
$0.65 \pm 0.02$.

\begin{figure}[htbp]
\centering
\begin{subfigure}[b]{0.47\textwidth}
\includegraphics[width=\linewidth]{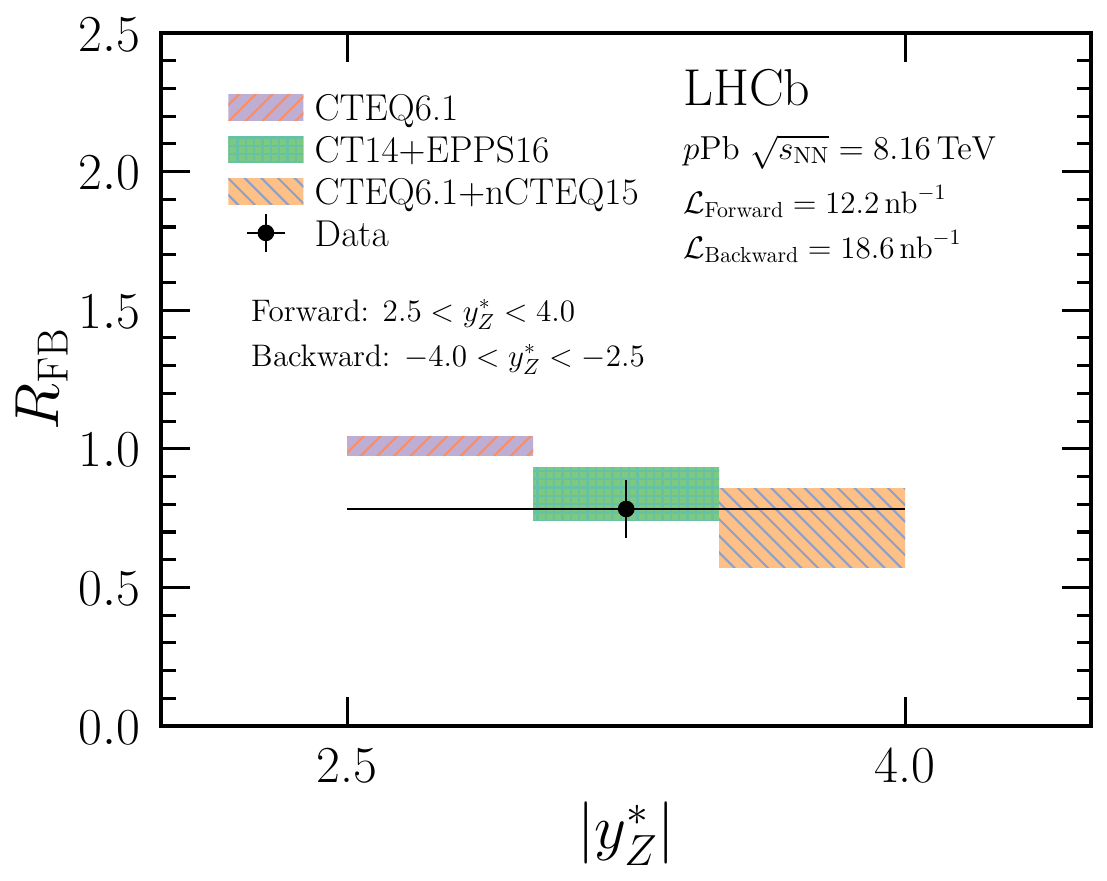}
\vspace*{0.0cm}
\end{subfigure}
\put(-40,105){(a)}
\hspace*{0.3cm}
\begin{subfigure}[b]{0.48\textwidth}
\includegraphics[width=\linewidth]{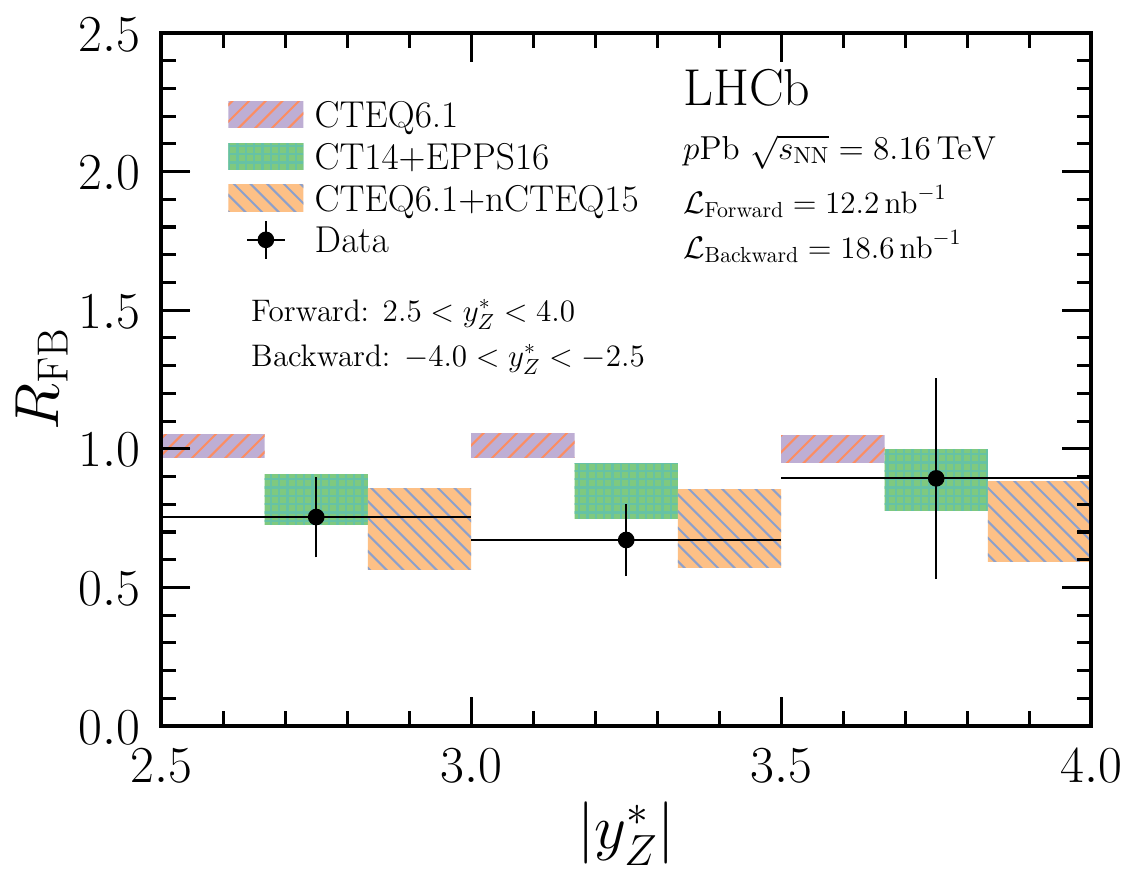}
\vspace*{0.05cm}
\end{subfigure}
\put(-40,100){(b)}
\\
\begin{subfigure}[b]{0.47\textwidth}
\includegraphics[width=\linewidth]{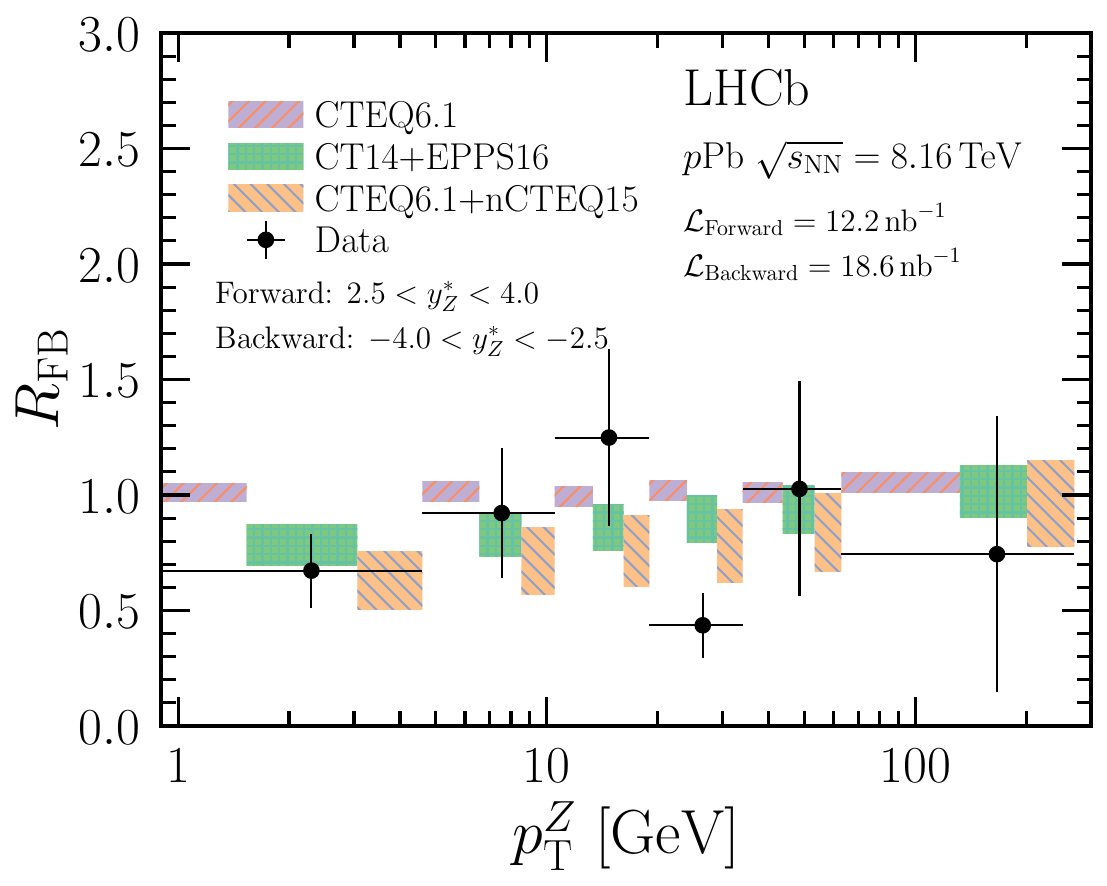}
\vspace*{0.0cm}
\end{subfigure}
\put(-40,115){(c)}
\hspace*{0.5cm}
\begin{subfigure}[b]{0.488\textwidth}
\includegraphics[width=\linewidth]{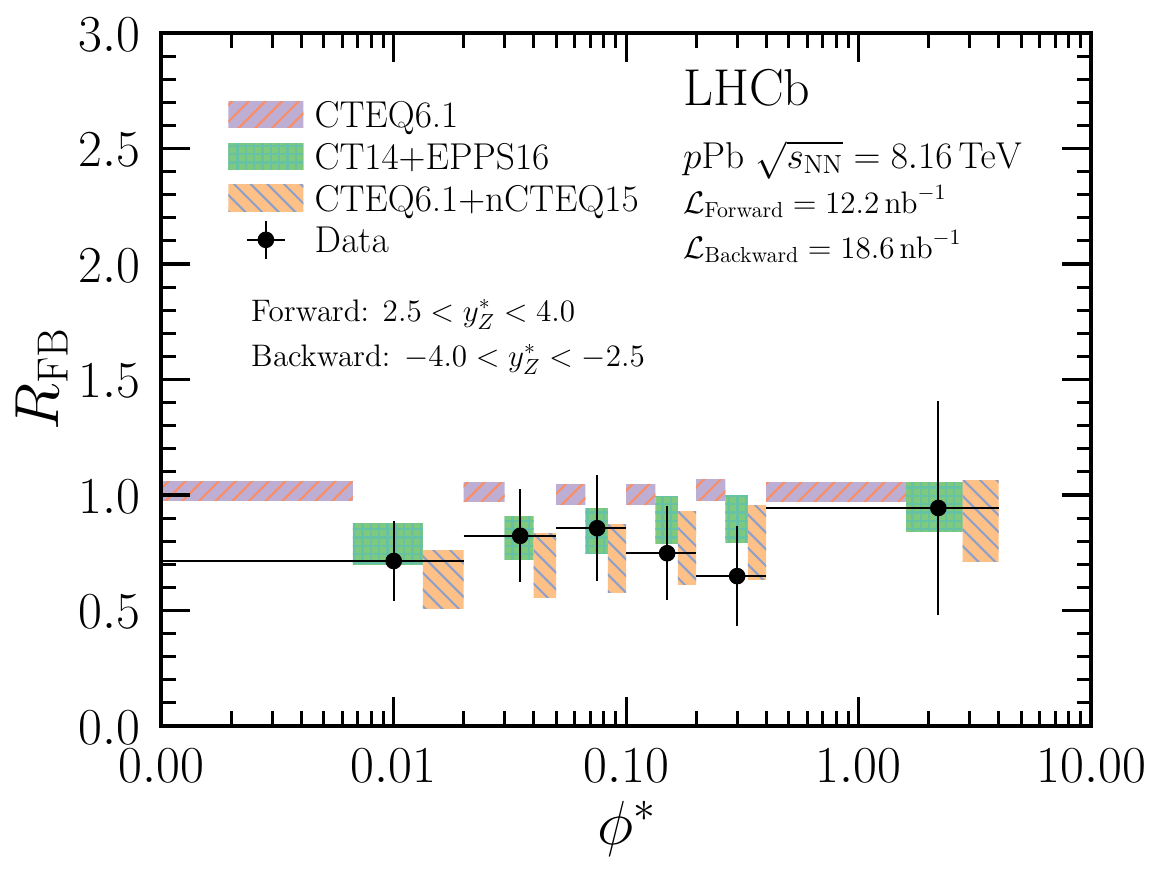}
\vspace*{0.15cm}
\end{subfigure}
\put(-45,115){(d)}
\caption{
The measured forward-backward ratio (\rfb),
(a) for the overall measurement,
(b) as a function of \zrapstar,
(c) as a function of \ZpT, and
(d) as a function of \phistar,
together with the \powheg predictions using CTEQ6.1, EPPS16 and nCTEQ15 (n)PDF sets.
}
\label{fig:result_RFB}
\end{figure}

The value of \rfb is then measured to be
$\rfb = 0.78 \pm 0.10$, 
as shown in Figure~\ref{fig:result_RFB}.
The measured \rfb value is below unity,
which is a reflection of the suppression
due to, \eg, nuclear shadowing at small
Bjorken-$x$, together with an average
enhancement at large Bjorken-$x$.
The data is in agreement
with the EPPS16 and nCTEQ15 predictions.
The uncertainty of the measurement is
smaller than the theoretical uncertainties
using EPPS16 and nCTEQ15 nPDFs,
which shows a constraining power on the
nPDFs.

In addition, the value of 
\rfb is also measured differentially as a function of
\zrapstar, \ZpT and \phistar.
The corresponding \kfb correction factors for the differential measurements
are derived in the same way
in separate intervals
as shown in Table~\ref{tab:kFB-diff}.
Figure~\ref{fig:result_RFB} (b), (c), and (d)
show the above measured values of \rfb as a function of
\zrapstar, \ZpT and \phistar, respectively,
together with the theory calculations.
The $\rfb$ measurement as a function of \zrapstar 
shows a general suppression below unity for all 
three intervals,
in good agreement
with the nPDF predictions.
For the two intervals in $2.5 < \zrapstar < 3.5$
the measurements give a smaller
uncertainty compared to the nCTEQ15 prediction
which can be used to constrain the
nPDFs.
For the $\rfb$ measurements as a function of \ZpT and \phistar,
a general suppression
is expected, however, 
the significance of the suppression is weak
given the large statistical fluctuation in the 
differential measurements.
From theory a
slightly stronger suppression at 
lower \ZpT or \phistar is expected,
but given the precision of the current 
measurement this tendency is not visible.
The \rfb as a function of \phistar shows 
a more stable trend than as a function of 
\ZpT, 
which exhibits a larger fluctuation
The detailed numerical values and uncertainties 
of the differential measurements of 
the \rfb together with the corresponding 
\fsrcorr factors are given in 
Appendix~\ref{apdx:cross-section}.

\begin{table}[htbp]
\begin{center}
\caption{The \murapstar acceptance correction factors (\kfb) for
\rfb measured in intervals of \zrapstar, \ZpT, and \phistar.}
\begin{footnotesize}
\begin{tabular}{rcccccc}
\hline\hline
\zrapstar & $[2.5, 3.0]$ & $[3.0, 3.5]$ & $[3.5,4.0]$ & & &\\ 
$\kfb(\zrapstar)$  & $0.29\pm0.01$ & $0.94\pm0.04$ & $2.73\pm0.13$  & & &\\
\hline
\ZpT \unitgev  & $[0, 4.6]$ & $[4.6, 10.5]$ & $[10.5, 19]$ & $[19, 34]$ & $[34,63]$ & $[63, 270]$ \\
$ \kfb(\ZpT) $ & $0.72\pm0.03$ & $0.65\pm0.03 $ & $0.63\pm0.03$ & $0.59\pm0.03$ & $0.58\pm0.02$ & $0.46\pm0.02$\\
\hline
\phistar  & $[0, 0.02]$ & $[0.02, 0.05]$ & $[0.05, 0.1]$ & $[0.1, 0.2]$ & $[0.2, 0.4]$ & $[0.4, 4]$ \\
$ \kfb(\phistar) $  & $0.70\pm0.03$ & $0.67\pm0.03$ & $0.65\pm0.03$ & $0.62\pm0.03$ & $0.59\pm0.03$ & $0.49\pm0.02$\\
\hline\hline
\end{tabular}
\end{footnotesize}
\label{tab:kFB-diff}
\end{center}
\end{table}

For the total nuclear modification factor measurements,
because the \zrapstar coverage between $pp$ and \pPb collisions
is different,
the
common \zrapstar acceptance range 
\mbox{$2.0<\zrapstar<4.0$} for forward rapidity
and
\mbox{$-4.0<\zrapstar<-2.5$} for backward rapidity is used.
The \murapstar acceptance correction factors
are determined to be
\mbox{$\kpa^{\rm fw.}=0.706\pm 0.002$}
and
\mbox{$\kpa^{\rm bw.}=1.518\pm 0.003$}
for forward and backward rapidities, respectively.
The cross-sections to be used in the \rpa
calculations are measured to be 
\mbox{$24.2\pm1.9\nb$}
and 
\mbox{$13.4\pm1.2\nb$}
for forward and backward collisions, respectively, 
in the corresponding \zrapstar common acceptance ranges.
The variables required to
calculate these cross-sections are shown in
Table~\ref{tab:system-uncertainties-sources}.
The corresponding $pp$ reference cross-sections at $8.16\tev$
are interpolated in the corresponding \zrapstar common acceptance ranges as 
\mbox{$(95.18\pm0.55)\times 10^{-3}\nb$}
and
\mbox{$(79.10\pm0.51)\times 10^{-3}\nb$}
for forward and backward rapidities, respectively.

\begin{figure}[tbph]
\begin{center}
\includegraphics[width=0.78\linewidth]{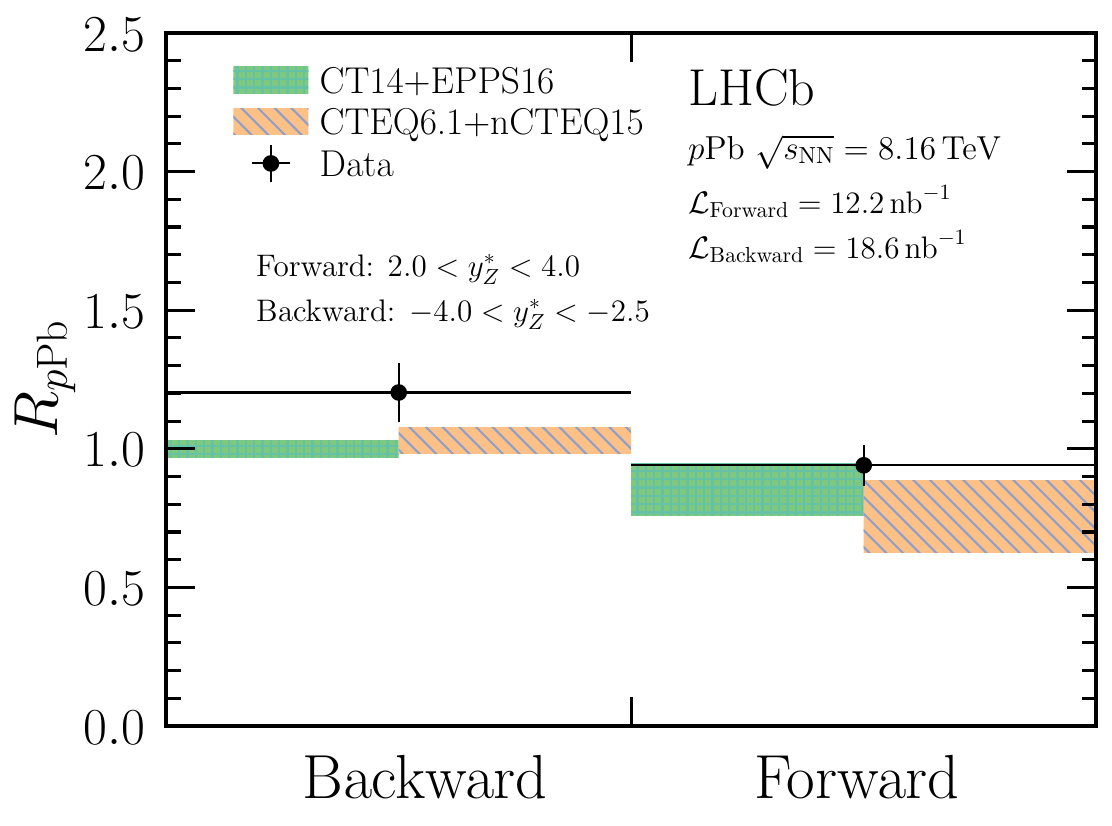}
\hspace*{0.5cm}
\end{center}
\caption{
Measurements of the nuclear modification factor (\rpa) compared to the \powheg predictions using the EPPS16 and nCTEQ15 (n)PDF sets.}
\label{fig:result_RpA}\vspace*{-0.5cm}
\end{figure}

The measured overall nuclear modification factors are
\mbox{$\rpa^{\rm fw}= 0.94\pm0.07$}
and
\mbox{$\rpa^{\rm bw}= 1.20\pm0.11$}
for forward and backward rapidities,
respectively,
where the total uncertainties are given.
These results are shown in Figure~\ref{fig:result_RpA}
compared with the \powheg predictions using the
EPPS16 and nCTEQ15 nPDF sets.
The overall \rpa results show good compatibility
between measurements and theoretical predictions.
The backward rapidity result shows
larger uncertainty compared to that of the
nPDF sets.
The measured central value is consistent with the
prediction at a 2$\sigma$ level.
The forward rapidity result
gives a higher precision than the
EPPS16 and nCTEQ15 nPDF sets,
and the central value is larger than
the prediction,
which shows a constraining power
on the current nPDF sets.

\begin{figure}[htbp]
\begin{center}
\vspace*{0.5cm}
\includegraphics[width=0.48\linewidth]{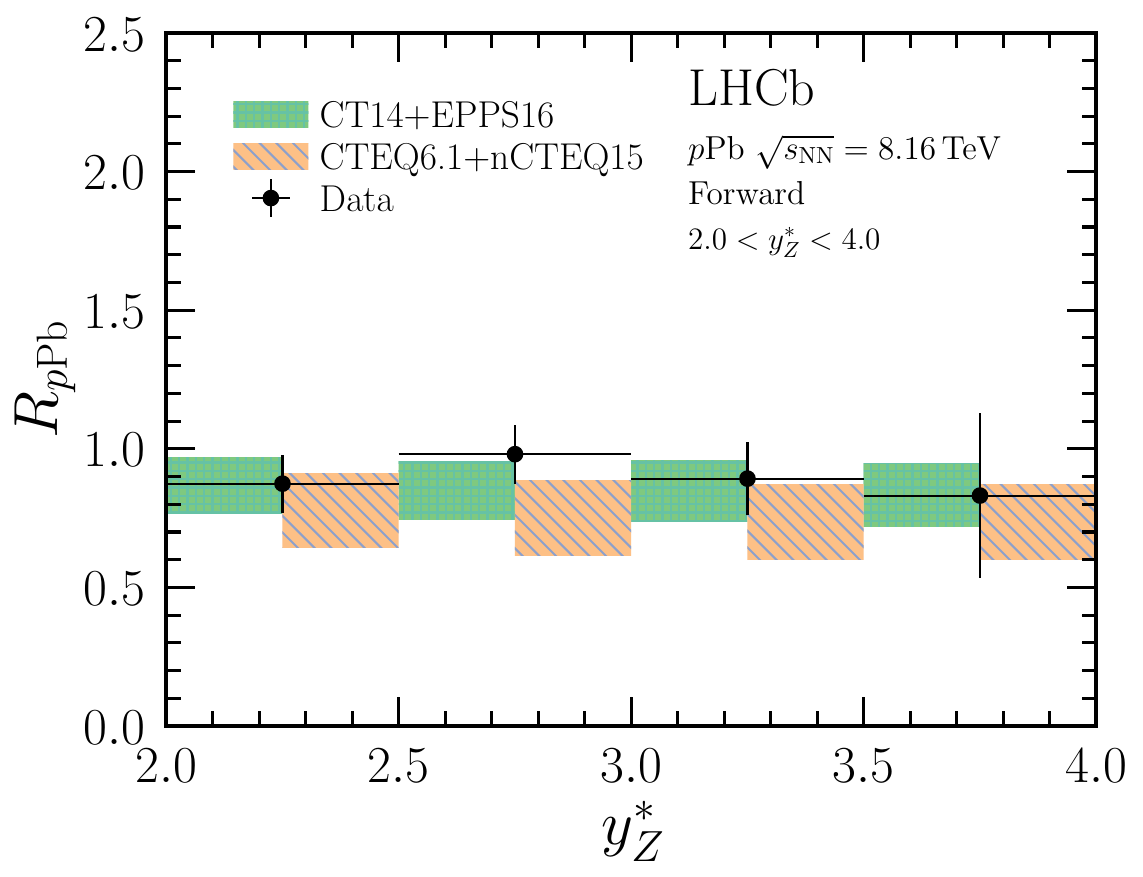}
\put(-170,45){(a)}
\hspace*{0.5cm}
\includegraphics[width=0.48\linewidth]{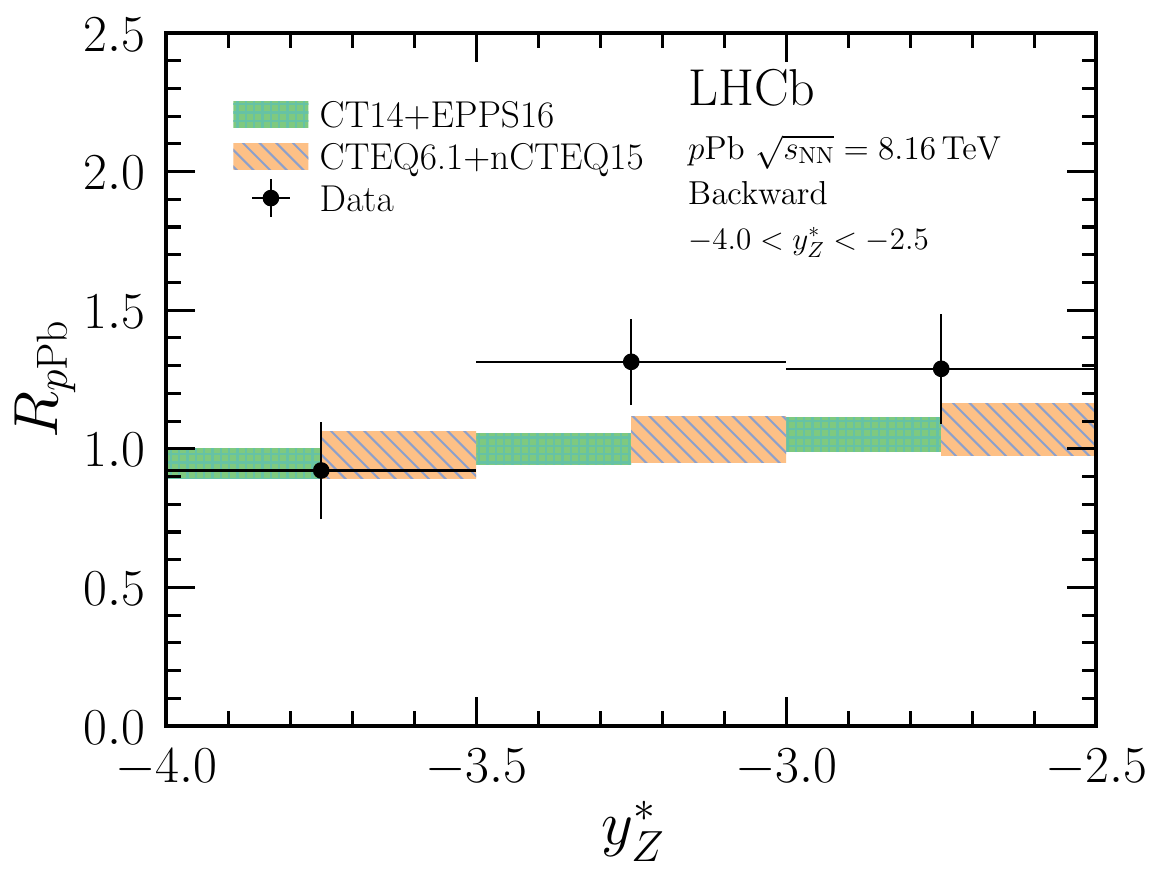}
\put(-170,45){(b)}
\\
\vspace*{0.5cm}
\hspace*{-0.2cm}
\includegraphics[width=0.455\linewidth]{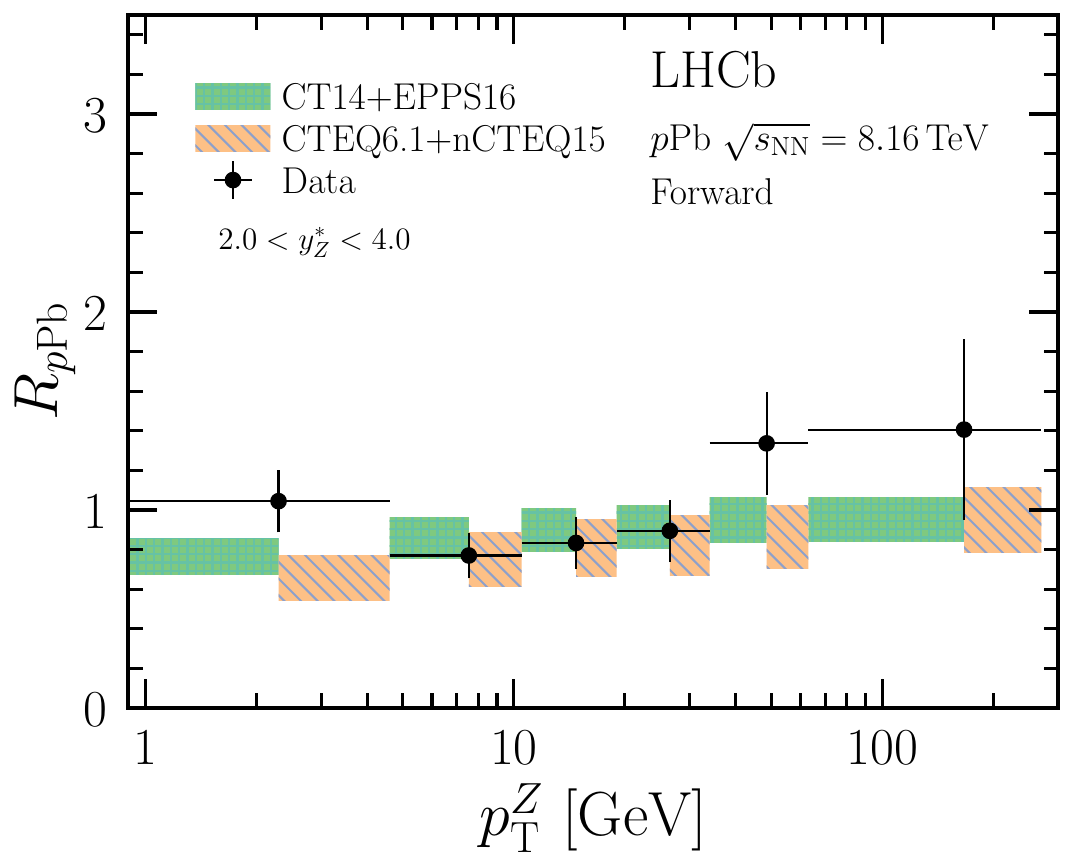}
\put(-170,45){(c)}
\hspace*{0.8cm}
\includegraphics[width=0.451\linewidth]{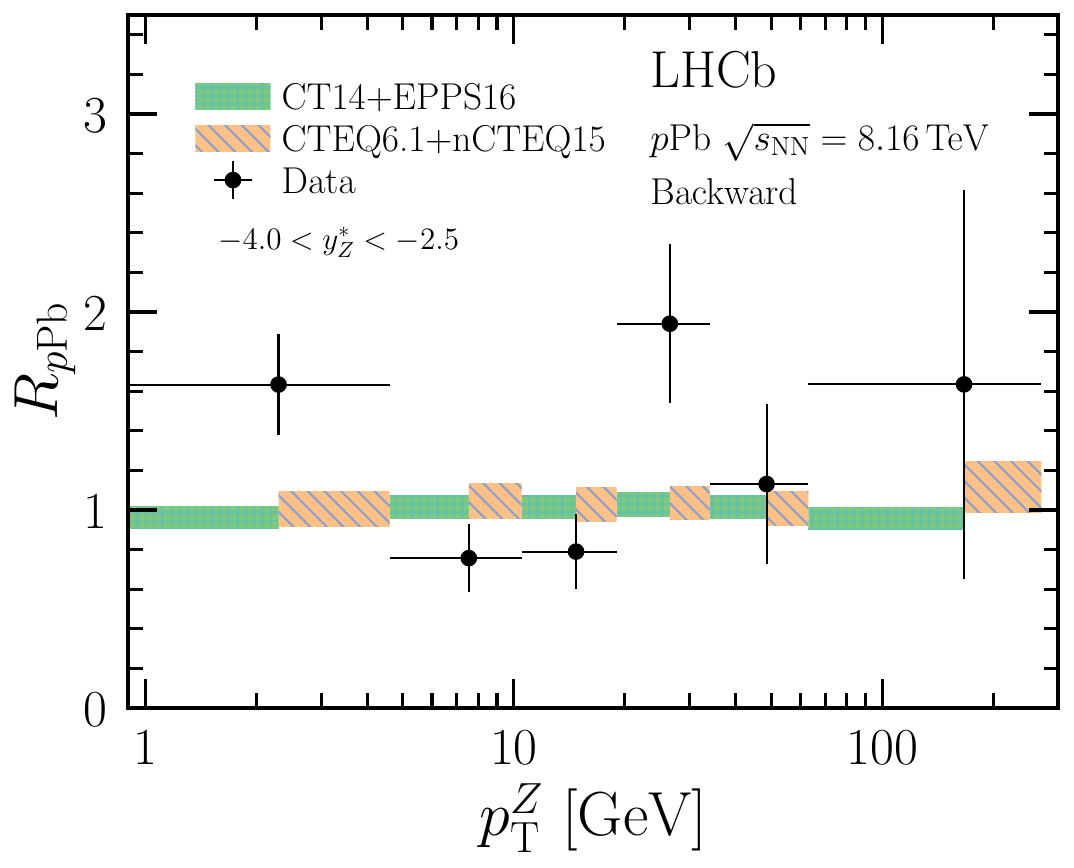}
\put(-170,45){(d)}
\\
\vspace*{0.5cm}
\includegraphics[width=0.485\linewidth]{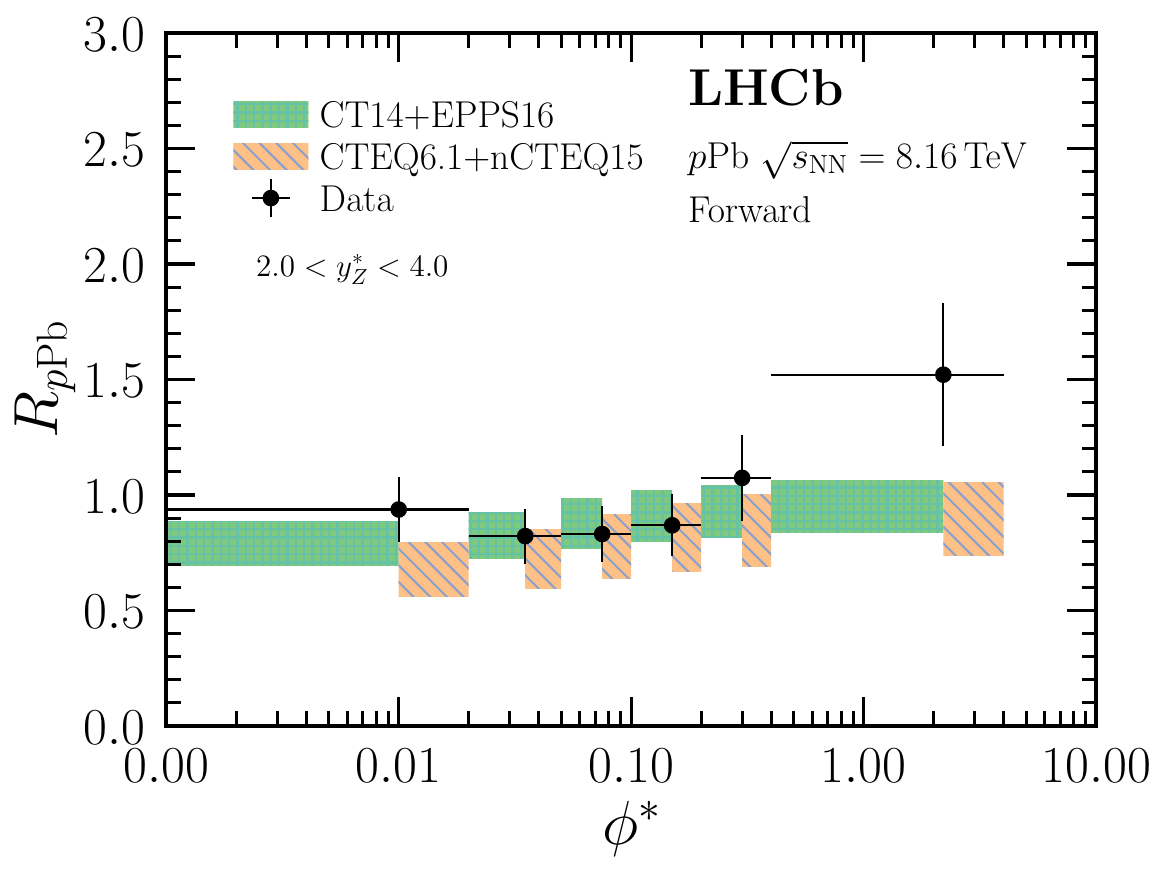}
\put(-170,45){(e)}
\hspace*{0.4cm}
\includegraphics[width=0.485\linewidth]{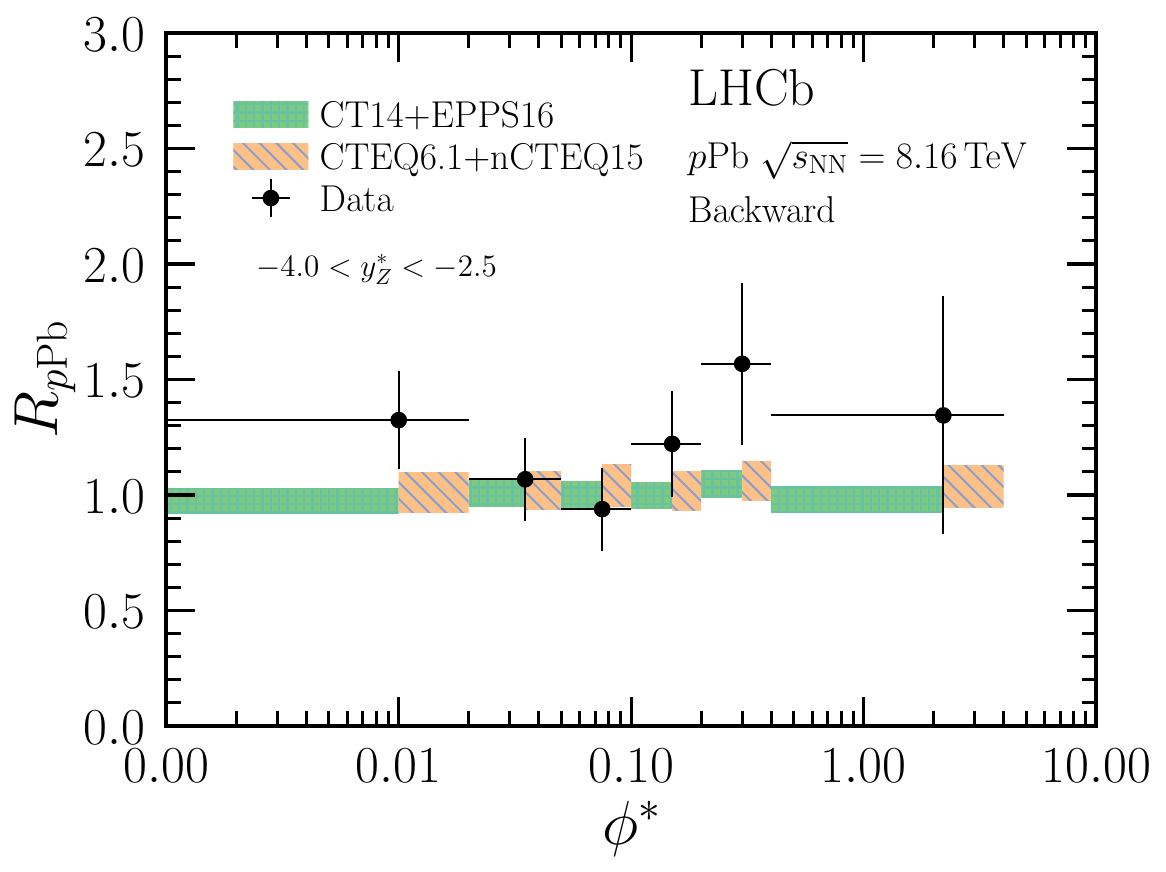}
\put(-170,45){(f)}
\end{center}
\caption{
Nuclear modification factors (\rpa) as a function of (top row) \zrapstar, (middle row) \ZpT and (bottom row) \phistar,
together with the \powheg prediction using
EPPS16 and nCTEQ15 nPDF sets,
where the left column is for forward collisions and the right column is for backward collisions.
}
\label{fig:rpA-bins}
\end{figure}

For the differential measurements of the \rpa,
the forward and backward cross-sections and the corresponding $pp$ reference cross-sections
are determined in the same way as for the total \rpa measurement but in intervals of
\zrapstar, \ZpT and \phistar.
The \murapstar acceptance correction factors \kpa for the differential measurements 
are also derived using \powheg
with the CTEQ6.1 PDF set considering theory uncertainties from
PDF variations and the factorisation and renormalisation scales,
as shown in Table~\ref{tab:kpA-diff}.
The resulting \rpa values as a function of \zrapstar,
\ZpT and \phistar are shown
in Figure~\ref{fig:rpA-bins}.
In general these results are compatible 
with nPDF predictions.
For backward rapidity, larger uncertainties
compared to the current nPDF predictions appear for
all three observables.
However, for forward rapidity
the larger dataset gives a higher precision for certain intervals
compared to the nPDF predictions.
The detailed numerical values and uncertainties of the \rpa as a function of
\zrapstar, \ZpT, and \phistar, together with the corresponding \fsrcorr factors are given in 
Appendix~\ref{apdx:cross-section}.

\begin{table}[htbp]
\begin{center}
\caption{The \murapstar acceptance correction factors (\kpa) for \rpa measured in intervals of \zrapstar, \ZpT, and \phistar for forward and backward collisions.}
\begin{footnotesize}
\begin{tabular}{rrcccccc}
\hline\hline
&\zrapstar & $[2.0, 2.5]$ & $[2.5, 3.0]$ & $[3.0, 3.5]$ & $[3.5,4.0]$ & & \\ 
\kpa(\zrapstar)& forward  & $0.3713$ & $0.7468$ & $1.2482$ & $2.1875$ & &\\
               &     & $\pm0.0005$ & $\pm0.0004$ & $\pm0.0010$ & $\pm0.0004$ & &\\\
& \zrapstar & $[-4.0, -3.5]$ & $[-3.5, -3.0]$ & $[-3.0, -2.5]$ & & & \\
\kpa(\zrapstar)& backward  & $0.7935$ & $1.3177$ & $2.5525$ & & & \\
               &      & $\pm0.0001$ & $\pm0.0001$ & $\pm0.0009$ & & & \\
\hline
&\ZpT \unitgev  & $[0, 4.6]$ & $[4.6, 10.5]$ & $[10.5, 19]$ & $[19, 34]$ & $[34,63]$ & $[63, 270]$ \\
\kpa(\ZpT) & forward  & $0.7821$ & $0.7212$ & $0.6903$ & $0.6613$ & $0.6136$ & $0.5455$ \\
           &          & $\pm0.0056$ & $\pm0.0049 $ & $\pm0.0045$ & $\pm0.0040$ & $\pm0.0035$ & $\pm0.0027$\\
           & backward & $1.6073$ & $1.7735$ & $1.8341$ & $1.9387$ & $2.0742$ & $2.8031$ \\
           &          & $\pm0.0034$ & $\pm0.0039$ & $\pm0.0040$ & $\pm0.0045$ & $\pm0.0047$ & $\pm 0.0069$\\
\hline
& \phistar  & $[0, 0.02]$ & $[0.02, 0.05]$ & $[0.05, 0.1]$ & $[0.1, 0.2]$ & $[0.2, 0.4]$ & $[0.4, 4]$ \\
\kpa(\phistar)& forward  & $0.7702$ & $0.7322$ & $0.7080$ & $0.6804$ & $0.6461$ & $0.5756$ \\
              &        & $\pm0.0054$ & $\pm0.0049$ & $\pm0.0047$ & $\pm0.0043$ & $\pm0.0040$ & $\pm 0.0032$\\
            & backward & $1.6400$ & $1.7113$ & $1.8105$ & $1.8645$ & $1.9741$ & $2.4238$ \\
            &          & $\pm0.0036$ & $\pm0.0036$ & $\pm0.0041$ & $\pm0.043$ & $\pm0.0047$ & $\pm0.0065$\\
\hline\hline
\end{tabular}
\end{footnotesize}
\label{tab:kpA-diff}
\end{center}
\end{table}

\section{Conclusion}

This article presents a comprehensive analysis of \Zmumu production
using \pPb collisions at 8.16\tev in the 
forward region recorded with the \lhcb detector.
The \Z-boson production fiducial cross-section, \rfb and \rpa are measured
inclusively and differentially as a 
function of \zrapstar, \ZpT, and \phistar.
The results are compatible with theoretical predictions from the
EPPS16 and nCTEQ15 nPDFs.
The precision of the measurements is significantly 
improved compared to the previous 
\lhcb results using  
\pPb collisions at a centre-of-mass energy of 5.02\tev~\cite{LHCb-PAPER-2014-022} collected in 2013. 
The measurements provide strong constraints
on the nPDFs,
 especially at high rapidity corresponding to the small Bjorken-$x$ region from $10^{-4}$ to $10^{-3}$.

\section*{Acknowledgements}
%
%
\noindent We express our gratitude to our colleagues in the CERN
accelerator departments for the excellent performance of the LHC. We
thank the technical and administrative staff at the LHCb
institutes.
We acknowledge support from CERN and from the national agencies:
CAPES, CNPq, FAPERJ and FINEP (Brazil); 
MOST and NSFC (China); 
CNRS/IN2P3 (France); 
BMBF, DFG and MPG (Germany); 
INFN (Italy); 
NWO (Netherlands); 
MNiSW and NCN (Poland); 
MEN/IFA (Romania); 
MICINN (Spain); 
SNSF and SER (Switzerland); 
NASU (Ukraine); 
STFC (United Kingdom); 
DOE NP and NSF (USA).
We acknowledge the computing resources that are provided by CERN, IN2P3
(France), KIT and DESY (Germany), INFN (Italy), SURF (Netherlands),
PIC (Spain), GridPP (United Kingdom), 
CSCS (Switzerland), IFIN-HH (Romania), CBPF (Brazil),
Polish WLCG  (Poland) and NERSC (USA).
We are indebted to the communities behind the multiple open-source
software packages on which we depend.
Individual groups or members have received support from
ARC and ARDC (Australia);
Minciencias (Colombia);
AvH Foundation (Germany);
EPLANET, Marie Sk\l{}odowska-Curie Actions and ERC (European Union);
A*MIDEX, ANR, IPhU and Labex P2IO, and R\'{e}gion Auvergne-Rh\^{o}ne-Alpes (France);
Key Research Program of Frontier Sciences of CAS, CAS PIFI, CAS CCEPP, 
Fundamental Research Funds for the Central Universities, Sci. \& Tech. Program of Guangzhou, and Guangdong Basic and Applied Basic Research Foundation (China);
GVA, XuntaGal, GENCAT and Prog.~Atracci\'on Talento, CM (Spain);
SRC (Sweden);
the Leverhulme Trust, the Royal Society
 and UKRI (United Kingdom).

\clearpage

\section*{Appendices}

\appendix

\section{Statistical and systematic correlation matrices}
\label{apdx:corr_matrix}

\begin{table}[htbp]
\begin{center}
\caption{Correlation matrix of statistical uncertainty in \zrapstar intervals for \forw collisions.}
\begin{tabular}{rrccccc}
\hline\hline
\zrapstar &Index  &1 &2 &3 &4 &5  \\
\hline
$[1.5, 2.0]$&1 &1.00 & & & & \\
$[2.0, 2.5]$&2 &0.01 &1.00 & & & \\
$[2.5, 3.0]$&3 &0.00 &0.01 &1.00 & & \\
$[3.0, 3.5]$&4 &0.00 &0.00 &0.01 &1.00 & \\
$[3.5, 4.0]$&5 &0.00 &0.00 &0.00 &0.00 &1.00 \\\hline\hline
\end{tabular}
\label{tab:stat_corr_matrix_zrap_pA_new}
\end{center}
\end{table}

\begin{table}[htbp]
\begin{center}
\caption{Correlation matrix of statistical uncertainty in \zrapstar intervals for \backw collisions.}
\begin{tabular}{rrccc}
\hline\hline
\zrapstar &Index &1 &2 &3 \\\hline
$[-3.0, -2.5]$&1 &1.00 & & \\
$[-3.5, -3.0]$&2 &0.01 &1.00 & \\
$[-4.0, -3.5]$&3 &0.00 &0.01 &1.00 \\\hline\hline
\end{tabular}
\label{tab:stat_corr_matrix_zrap_Ap_new}
\end{center}
\end{table}

\begin{table}[htbp]
\begin{center}
\caption{Correlation matrix of statistical uncertainty in \ZpT intervals for \forw collisions.}
\begin{tabular}{rrcccccc}
\hline\hline
\ZpT \unitgev &Index&1 &2 &3 &4 &5 &6 \\
\hline
$[0.0,   4.6]$&1 &1.00 & & & & & \\
$[4.6,  10.5]$&2 &0.93 &1.00 & & & & \\
$[10.5, 19.0]$&3 &0.00 &0.05 &1.00 & & & \\
$[19.0, 34.0]$&4 &0.00 &0.00 &0.04 &1.00 & & \\
$[34.0, 63.0]$&5 &0.00 &0.00 &0.00 &0.03 &1.00 & \\
$[63.0,270.0]$&6 &0.00 &0.00 &0.00 &0.00 &0.02 &1.00 \\\hline\hline
\end{tabular}
\label{tab:stat_corr_matrix_zpt_pA_new}
\end{center}
\end{table}

\begin{table}[htbp]
\begin{center}
\caption{Correlation matrix of statistical uncertainty in \ZpT intervals for \backw collisions.}
\begin{tabular}{rrcccccc}
\hline\hline
\ZpT \unitgev &Index &1 &2 &3 &4 &5 &6 \\
\hline
$[0.0,   4.6]$&1  &1.00 & & & & & \\
$[4.6,  10.5]$&2  &0.08 &1.00 & & & & \\
$[10.5, 19.0]$&3 &0.00 &0.04 &1.00 & & & \\
$[19.0, 34.0]$&4 &0.00 &0.00 &0.03 &1.00 & & \\
$[34.0, 63.0]$&5 &0.00 &0.00 &0.00 &0.02 &1.00 & \\
$[63.0,270.0]$&6 &0.00 &0.00 &0.00 &0.00 &0.01 &1.00 \\\hline\hline
\end{tabular}
\label{tab:stat_corr_matrix_zpt_Ap_new}
\end{center}
\end{table}

\begin{table}[htbp]
\begin{center}
\caption{Correlation matrix of statistical uncertainty in \phistar intervals for \forw collisions.}
\begin{tabular}{rrcccccc}
\hline\hline
\phistar &Index &1 &2 &3 &4 &5 &6 \\
\hline
$[0.00, 0.02]$&1 &1.00 & & & & & \\
$[0.02, 0.05]$&2 &0.01 &1.00 & & & & \\
$[0.05, 0.10]$&3 &0.00 &0.00 &1.00 & & & \\
$[0.10, 0.20]$&4 &0.00 &0.00 &0.00 &1.00 & & \\
$[0.20, 0.40]$&5 &0.00 &0.00 &0.00 &0.00 &1.00 & \\
$[0.40, 4.00]$&6 &0.00 &0.00 &0.00 &0.00 &0.00 &1.00 \\\hline\hline
\end{tabular}
\label{tab:stat_corr_matrix_zps_pA_new}
\end{center}
\end{table}

\begin{table}[htbp]
\begin{center}
\caption{Correlation matrix of statistical uncertainty in \phistar intervals for \backw collisions.}
\begin{tabular}{rrcccccc}
\hline\hline
\phistar &Index &1 &2 &3 &4 &5 &6 \\
\hline
$[0.00, 0.02]$&1 &1.00 & & & & & \\
$[0.02, 0.05]$&2 &0.00 &1.00 & & & & \\
$[0.05, 0.10]$&3 &0.00 &0.00 &1.00 & & & \\
$[0.10, 0.20]$&4 &0.00 &0.00 &0.00 &1.00 & & \\
$[0.20, 0.40]$&5 &0.00 &0.00 &0.00 &0.00 &1.00 & \\
$[0.40, 4.00]$&6 &0.00 &0.00 &0.00 &0.00 &0.00 &1.00 \\\hline\hline
\end{tabular}
\label{tab:stat_corr_matrix_zps_Ap)new}
\end{center}
\end{table}

\begin{table}[htbp]
\begin{center}
\caption{Correlation matrix of statistical uncertainty in \ZpT intervals for \forw collisions in the low \ZpT range.}
\begin{tabular}{rrcccccccc}
\hline\hline
\ZpT \unitgev &Index &1 &2 &3 &4 &5 &6 &7 &8  \\
\hline
$[0.0, 2.2]$&1 &1.00 & & & & & & & \\
$[2.2, 4.6]$&2 &0.24 &1.00 & & & & & &  \\
$[4.6, 7.2]$&3 &0.04 &0.26 &1.00 & & & & & \\
$[7.2, 8.7]$&4 &0.01 &0.04 &0.25 &1.00 & & & &  \\
$[8.7,10.5]$&5 &0.01 &0.01 &0.04 &0.24 &1.00 & & & \\
$[10.5,12.8]$&6 &0.00 &0.01 &0.01 &0.03 &0.21 &1.00 & & \\
$[12.8,15.4]$&7 &0.00 &0.00 &0.01 &0.01 &0.03 &0.20 &1.00 & \\
$[15.4, 19.0]$&8 &0.00 &0.00 &0.00 &0.01 &0.01 &0.02 &0.18 &1.00  \\\hline\hline
\end{tabular}
\label{tab:stat_corr_matrix_zpt_pA}
\end{center}
\end{table}

\begin{table}[htbp]
\begin{center}
\caption{Correlation matrix of statistical uncertainty in \ZpT intervals for \backw collisions in the low \ZpT range.}
\begin{tabular}{rrcccccccc}
\hline\hline
\ZpT \unitgev &Index  &1 &2 &3 &4 &5 &6 &7 &8\\
\hline
$[0.0, 2.2]$&1 &1.00 & & & & & & &  \\
$[2.2, 4.6]$&2 &0.20 &1.00 & & & & & &  \\
$[4.6, 7.2]$&3 &0.03 &0.22 &1.00 & & & & &  \\
$[7.2, 8.7]$&4 &0.01 &0.03 &0.21 &1.00 & & & & \\
$[8.7,10.5]$&5 &0.01 &0.01 &0.02 &0.20 &1.00 & & & \\
$[10.5,12.8]$&6 &0.00 &0.00 &0.01 &0.02 &0.18 &1.00 & & \\
$[12.8, 15.4]$&7 &0.00 &0.00 &0.00 &0.01 &0.02 &0.17 &1.00 &  \\
$[15.4, 19.0]$&8 &0.00 &0.00 &0.00 &0.00 &0.01 &0.02 &0.14 &1.00  \\\hline\hline
\end{tabular}
\label{tab:stat_corr_matrix_zpt_Ap}
\end{center}
\end{table}

\begin{table}[htbp]
\begin{center}
\caption{Correlation matrix of efficiency uncertainty in \zrapstar intervals for \forw collisions.}
\begin{tabular}{rrccccc}
\hline\hline
\zrapstar &Index &1 &2 &3 &4 &5  \\
\hline
$[2.0, 2.5]$&1 &1.00 & & & & \\
$[2.5, 3.0]$&2 &0.71 &1.00 & & & \\
$[3.0, 3.5]$&3 &0.37 &0.72 &1.00 & & \\
$[3.5, 4.0]$&4 &0.03 &0.24 &0.71 &1.00 & \\
$[4.0, 4.5]$&5 &0.00 &0.05 &0.45 &0.79 &1.00 \\\hline\hline
\end{tabular}
\label{tab:syst_corr_matrix_zrap_pA}
\end{center}
\end{table}

\begin{table}[htbp]
\begin{center}
\caption{Correlation matrix of efficiency uncertainty in \zrapstar intervals for \backw collisions.}
\begin{tabular}{rrccc}
\hline\hline
\zrapstar &Index &1 &2 &3 \\\hline
$[-3.0, -2.5]$&1 &1.00 & & \\
$[-3.5, -3.0]$&2 &0.64 &1.00 & \\
$[-4.0, -3.5]$&3 &0.48 &0.75 &1.00 \\\hline\hline
\end{tabular}
\label{tab:syst_corr_matrix_zrap_Ap}
\end{center}
\end{table}

\begin{table}[htbp]
\begin{center}
\caption{Correlation matrix of efficiency uncertainty in \ZpT intervals for \forw collisions.}
\begin{tabular}{rrcccccc}
\hline\hline
\ZpT \unitgev &Index &1 &2 &3 &4 &5 &6 \\
\hline
$[0.0, 4.6]$&  1 &1.00 & & & & & \\
$[4.6, 10.5]$& 2 &0.88 &1.00 & & & & \\
$[10.5, 19.0]$&3 &0.87 &0.89 &1.00 & & & \\
$[19.0, 34.0]$&4 &0.92 &0.89 &0.91 &1.00 & & \\
$[34.0, 63.0]$&5 &0.85 &0.80 &0.79 &0.87 &1.00 & \\
$[63.0,270.0]$&6 &0.75 &0.82 &0.82 &0.82 &0.74 &1.00 \\\hline\hline
\end{tabular}
\label{tab:syst_corr_matrix_zpt_pA}
\end{center}
\end{table}

\begin{table}[htbp]
\begin{center}
\caption{Correlation matrix of efficiency uncertainty in \ZpT intervals for \backw collisions.}
\begin{tabular}{rrcccccc}
\hline\hline
\ZpT \unitgev &Index &1 &2 &3 &4 &5 &6 \\
\hline
$[0.0, 4.6]$&  1 &1.00 & & & & & \\
$[4.6, 10.5]$& 2 &0.85 &1.00 & & & & \\
$[10.5, 19.0]$&3 &0.91 &0.91 &1.00 & & & \\
$[19.0, 34.0]$&4 &0.83 &0.94 &0.86 &1.00 & & \\
$[34.0, 63.0]$&5 &0.68 &0.84 &0.65 &0.80 &1.00 & \\
$[63.0,270.0]$&6 &0.50 &0.71 &0.65 &0.70 &0.77 &1.00 \\\hline\hline
\end{tabular}
\label{tab:syst_corr_matrix_zpt_Ap}
\end{center}
\end{table}

\begin{table}[htbp]
\begin{center}
\caption{Correlation matrix of efficiency uncertainty in \phistar intervals for \forw collisions.}
\begin{tabular}{rrcccccc}
\hline\hline
\phistar &Index &1 &2 &3 &4 &5 &6 \\
\hline
$[0.00, 0.02]$&1 &1.00 & & & & & \\
$[0.02, 0.05]$&2 &0.83 &1.00 & & & & \\
$[0.05, 0.10]$&3 &0.92 &0.90 &1.00 & & & \\
$[0.10, 0.20]$&4 &0.83 &0.85 &0.80 &1.00 & & \\
$[0.20, 0.40]$&5 &0.85 &0.71 &0.77 &0.73 &1.00 & \\
$[0.40, 4.00]$&6 &0.85 &0.89 &0.85 &0.81 &0.82 &1.00 \\\hline\hline
\end{tabular}
\label{tab:syst_corr_matrix_zps_pA}
\end{center}
\end{table}

\begin{table}[htbp]
\begin{center}
\caption{Correlation matrix of efficiency uncertainty in \phistar intervals for \backw collisions.}
\begin{tabular}{rrcccccc}
\hline\hline
\phistar&Index &1 &2 &3 &4 &5 &6 \\
\hline
$[0.00, 0.02]$&1 &1.00 & & & & & \\
$[0.02, 0.05]$&2 &0.90 &1.00 & & & & \\
$[0.05, 0.10]$&3 &0.88 &0.80 &1.00 & & & \\
$[0.10, 0.20]$&4 &0.80 &0.79 &0.77 &1.00 & & \\
$[0.20, 0.40]$&5 &0.76 &0.70 &0.79 &0.58 &1.00 & \\
$[0.40, 4.00]$&6 &0.53 &0.60 &0.55 &0.75 &0.54 &1.00 \\\hline\hline
\end{tabular}
\label{tab:syst_corr_matrix_zps_Ap}
\end{center}
\end{table}

\begin{table}[htbp]
\begin{center}
\caption{Correlation matrix of efficiency uncertainty in \ZpT intervals for \forw collisions in the low \ZpT range.}
\begin{footnotesize}
\begin{tabular}{rrcccccccc}
\hline\hline
\ZpT \unitgev&Index &1 &2 &3 &4 &5 &6 &7 &8  \\
\hline
$[0.0,2.2]$&  1 &1.00        \\
$[2.2,4.6]$&  2 &0.57        &1.00   \\
$[4.6,7.2]$&  3 &0.71        &0.83   &1.00   \\
$[7.2,8.7]$&  4 &0.57        &0.67   &0.68   &1.00   \\
$[8.7,10.5]$& 5 &0.61        &0.71   &0.80   &0.68   &1.00   \\
$[10.5,12.8]$&6 &0.72        &0.73   &0.83   &0.64   &0.85   &1.00   \\
$[12.8,15.4]$&7 &0.66        &0.77   &0.84   &0.62   &0.71   &0.85   &1.00   \\
$[15.4,19.0]$&8 &0.58        &0.78   &0.78   &0.46   &0.76   &0.77   &0.70   &1.00   \\\hline\hline
\end{tabular}
\end{footnotesize}
\label{tab:syst_corr_matrix_zpt_fine_pA}
\end{center}
\begin{center}
\caption{Correlation matrix of efficiency uncertainty in \ZpT intervals for \backw collisions in the low \ZpT range.}
\begin{footnotesize}
\begin{tabular}{rrcccccccc}
\hline\hline
\ZpT \unitgev &Index &1 &2 &3 &4 &5 &6 &7 &8  \\
\hline
$[0.0,2.2]$&  1 &1.00        \\
$[2.2,4.6]$&  2 &0.82        &1.00   \\
$[4.6,7.2]$&  3 &0.73        &0.76   &1.00   \\
$[7.2,8.7]$&  4 &0.78        &0.66   &0.70   &1.00   \\
$[8.7,10.5]$& 5 &0.48        &0.74   &0.58   &0.48   &1.00   \\
$[10.5,12.8]$&6 &0.86        &0.84   &0.69   &0.61   &0.58   &1.00   \\
$[12.8,15.4]$&7 &0.72        &0.72   &0.67   &0.88   &0.55   &0.52   &1.00   \\
$[15.4,19.0]$&8 &0.79        &0.78   &0.81   &0.81   &0.55   &0.74   &0.71   &1.00   \\\hline\hline
\end{tabular}
\end{footnotesize}
\label{tab:syst_corr_matrix_zpt_fine_Ap}
\end{center}
\end{table}

\clearpage

\section{Differential cross-section, forward-backward ratio, and nuclear modification factors}
\label{apdx:cross-section}

\begin{table}[htbp]
\begin{center}
\caption{ Differential cross-section of \Zmumu in intervals of \zrapstar for \forw and \backw collisions, together with the FSR correction.
In the differential cross-section results, the first uncertainty is statistical, the second is systematic and the third is from integrated luminosity.}
\begin{tabular}{rrrr}
\hline\hline
& \multicolumn{1}{c}{\zrapstar}     & \multicolumn{1}{c}{$d\sigma/d\zrapstar\,[\nb]$} & \multicolumn{1}{c}{\fsrcorr} \\ \hline
\For 
&$ [1.5,2.0]$    &       $ 5.60\pm 1.08\pm 0.20\pm 0.15$         &       $ 1.017\pm0.002$  \\
&$ [2.0,2.5]$    &       $ 16.19\pm 1.80\pm 0.53\pm 0.43$        &       $ 1.020\pm0.001$  \\
&$ [2.5,3.0]$    &       $ 20.36\pm 2.02\pm 0.72\pm 0.53$        &       $ 1.023\pm0.001$  \\
&$ [3.0,3.5]$    &       $ 10.05\pm 1.42\pm 0.36\pm 0.26$        &       $ 1.030\pm0.001$  \\
&$ [3.5,4.0]$    &       $ 1.59\pm 0.56\pm 0.06\pm 0.04$         &       $ 1.014\pm0.002$  \\
\hline
\Back
&$ [-4.0,-3.5]$       &       $ 4.87\pm 0.89\pm 0.17\pm 0.18$        &       $ 1.028\pm0.002$  \\
&$ [-3.5,-3.0]$       &       $ 14.02\pm 1.49\pm 0.49\pm 0.53$       &       $ 1.018\pm0.001$  \\
&$ [-3.0,-2.5]$       &       $ 7.83\pm 1.13\pm 0.30\pm 0.30$        &       $ 1.020\pm0.001$  \\
\hline\hline
\end{tabular}\vspace*{0.5cm}
\label{tab:diffxsec-y}
\end{center}
\end{table}

\begin{table}[htbp]
\begin{center}
\caption{ Differential cross-section of \Zmumu in intervals of \ZpT for \forw and \backw collisions, together with the FSR correction.
In the differential cross-section results, the first uncertainty is statistical, the second is systematic and the third is from integrated luminosity.}
\begin{tabular}{rrrr}
\hline\hline
& \multicolumn{1}{c}{\ZpT\,\unitgev}     & \multicolumn{1}{c}{$d\sigma/d\ZpT\,[\nb/\gev]$} & \multicolumn{1}{c}{\fsrcorr} \\ \hline
\For 
&$ [0.0,4.6]$    &       $ 1.441\pm 0.209\pm 0.046\pm 0.038$        &       $ 1.068\pm0.002$  \\
&$ [4.6,10.5]$   &       $ 1.086\pm 0.156\pm 0.035\pm 0.029$        &       $ 1.033\pm0.001$  \\
&$ [10.5,19.0]$  &       $ 0.612\pm 0.092\pm 0.020\pm 0.016$        &       $ 0.994\pm0.002$  \\
&$ [19.0,34.0]$  &       $ 0.272\pm 0.046\pm 0.010\pm 0.007$        &       $ 0.988\pm0.002$  \\
&$ [34.0,63.0]$  &       $ 0.124\pm 0.023\pm 0.005\pm 0.003$        &       $ 1.034\pm0.002$  \\
&$ [63.0,270.0]$ &       $ (5.9\pm 1.9\pm 0.4\pm 0.2)\times 10^{-3}$&       $ 1.043\pm0.007$  \\
\hline
\Back 
&$ [0.0,4.6]$    &       $ 1.097\pm 0.162\pm 0.042\pm 0.027$        &       $ 1.082\pm0.002$  \\
&$ [4.6,10.5]$   &       $ 0.434\pm 0.096\pm 0.016\pm 0.011$        &       $ 1.025\pm0.002$  \\
&$ [10.5,19.0]$  &       $ 0.218\pm 0.051\pm 0.009\pm 0.005$        &       $ 0.976\pm0.002$  \\
&$ [19.0,34.0]$  &       $ 0.201\pm 0.040\pm 0.009\pm 0.005$        &       $ 0.981\pm0.002$  \\
&$ [34.0,63.0]$  &       $ 0.031\pm 0.011\pm 0.002\pm 0.001$        &       $ 1.028\pm0.003$  \\
&$ [63.0,270.0]$ &       $ (1.3\pm 0.8\pm 0.1\pm 0.1)\times 10^{-3}$&       $ 1.064\pm0.009$  \\
\hline\hline
\end{tabular}\vspace*{0.5cm}
\label{tab:diffxsec-zpt}
\end{center}
\end{table}

\begin{table}[htbp]
\begin{center}
\caption{ Differential cross-section of \Zmumu in intervals of \ZpT for \forw and \backw collisions in the low \ZpT region, together with the FSR correction.
In the differential cross-section results, the first uncertainty is statistical, the second is systematic and the third is from integrated luminosity.}
\begin{tabular}{rrrr}
\hline\hline
& \multicolumn{1}{c}{\ZpT\,\unitgev}     & \multicolumn{1}{c}{$d\sigma/d\ZpT\,[\nb/\gev]$} & \multicolumn{1}{c}{\fsrcorr} \\ \hline
\For 
&$ [0.0,2.2]$    &       $ 1.463\pm 0.352\pm 0.059\pm 0.038$        &       $ 1.079\pm0.002$  \\
&$ [2.2,4.6]$    &       $ 1.450\pm 0.347\pm 0.045\pm 0.038$        &       $ 1.064\pm0.002$  \\
&$ [4.6,7.2]$    &       $ 1.414\pm 0.310\pm 0.044\pm 0.037$        &       $ 1.048\pm0.004$  \\
&$ [7.2,8.7]$    &       $ 0.742\pm 0.344\pm 0.027\pm 0.019$        &       $ 1.035\pm0.004$  \\
&$ [8.7,10.5]$   &       $ 0.906\pm 0.261\pm 0.044\pm 0.024$        &       $ 1.005\pm0.002$  \\
&$ [10.5,12.8]$  &       $ 0.901\pm 0.227\pm 0.037\pm 0.024$        &       $ 1.003\pm0.003$  \\
&$ [12.8,15.4]$  &       $ 0.534\pm 0.167\pm 0.020\pm 0.014$        &       $ 1.000\pm0.004$  \\
&$ [15.4,19.0]$  &       $ 0.475\pm 0.130\pm 0.022\pm 0.012$        &       $ 0.981\pm0.002$  \\
\hline
\Back 
&$ [0.0,2.2]$    &       $ 0.813\pm 0.223\pm 0.029\pm 0.020$        &       $ 1.107\pm0.004$  \\
&$ [2.2,4.6]$    &       $ 1.352\pm 0.259\pm 0.062\pm 0.033$        &       $ 1.072\pm0.004$  \\
&$ [4.6,7.2]$    &       $ 0.596\pm 0.166\pm 0.024\pm 0.015$        &       $ 1.037\pm0.006$  \\
&$ [7.2,8.7]$    &       $ 0.438\pm 0.187\pm 0.024\pm 0.011$        &       $ 1.025\pm0.003$  \\
&$ [8.7,10.5]$   &       $ 0.220\pm 0.119\pm 0.011\pm 0.005$        &       $ 0.999\pm0.003$  \\
&$ [10.5,12.8]$  &       $ 0.232\pm 0.104\pm 0.008\pm 0.006$        &       $ 0.991\pm0.004$  \\
&$ [12.8,15.4]$  &       $ 0.168\pm 0.083\pm 0.008\pm 0.004$        &       $ 0.965\pm0.006$  \\
&$ [15.4,19.0]$  &       $ 0.250\pm 0.083\pm 0.012\pm 0.006$        &       $ 0.970\pm0.004$  \\
\hline\hline
\end{tabular}\vspace*{0.5cm}
\label{tab:diffxsec-zpt-lowpt}
\end{center}
\end{table}

\begin{table}[htbp]
\begin{center}
\caption{ Differential cross-section of \Zmumu in intervals of \phistar for \forw and \backw collisions, together with the FSR correction.
In the differential cross-section results, the first uncertainty is statistical, the second is systematic and the third is from integrated luminosity.}
\begin{tabular}{rrrr}
\hline\hline
& \multicolumn{1}{c}{\phistar}     & \multicolumn{1}{c}{$d\sigma/d\phistar\,[\nb]$} & \multicolumn{1}{c}{\fsrcorr} \\ \hline
\For 
&$ [0.0,0.0]$    &       $ 261.288\pm           36.234\pm           10.959\pm 6.865$        &       $ 1.031\pm0.003$  \\
&$ [0.0,0.1]$    &       $ 179.805\pm           24.698\pm \phantom{1}6.125\pm 4.724$        &       $ 1.044\pm0.002$  \\
&$ [0.1,0.1]$    &       $ 104.654\pm           14.375\pm \phantom{1}3.886\pm 2.750$        &       $ 1.010\pm0.002$  \\
&$ [0.1,0.2]$    &       $  46.938\pm \phantom{1}6.847\pm \phantom{1}1.765\pm 1.233$        &       $ 1.018\pm0.001$  \\
&$ [0.2,0.4]$    &       $  17.902\pm \phantom{1}2.984\pm \phantom{1}0.656\pm 0.470$        &       $ 1.004\pm0.002$  \\
&$ [0.4,4.0]$    &       $   0.786\pm \phantom{1}0.151\pm \phantom{1}0.042\pm 0.021$        &       $ 1.024\pm0.004$  \\
\hline
\Back 
&$ [0.0,0.0]$    &       $ 173.372\pm           26.752\pm \phantom{1}6.029\pm 4.292$        &       $ 1.053\pm0.003$  \\
&$ [0.0,0.1]$    &       $  99.962\pm           16.216\pm \phantom{1}3.547\pm 2.475$        &       $ 1.007\pm0.002$  \\
&$ [0.1,0.1]$    &       $  46.233\pm \phantom{1}8.585\pm \phantom{1}1.672\pm 1.145$        &       $ 1.015\pm0.002$  \\
&$ [0.1,0.2]$    &       $  24.080\pm \phantom{1}4.396\pm \phantom{1}0.863\pm 0.596$        &       $ 1.014\pm0.001$  \\
&$ [0.2,0.4]$    &       $   8.552\pm \phantom{1}1.866\pm \phantom{1}0.355\pm 0.212$        &       $ 1.015\pm0.002$  \\
&$ [0.4,4.0]$    &       $   0.165\pm \phantom{1}0.062\pm \phantom{1}0.009\pm 0.004$        &       $ 1.023\pm0.004$  \\
\hline\hline
\end{tabular}\vspace*{0.5cm}
\label{tab:diffxsec-zps}
\end{center}
\end{table}

\begin{table}[htbp]
\begin{center}
\caption{The forward-backward ratio (\rfb) in intervals of $|\zrapstar|$, together with the FSR correction.
For the \rfb results, the first uncertainty is statistical, the second is systematic and the third is from integrated luminosity.}
\begin{tabular}{rrccc}
\hline\hline
\multicolumn{1}{c}{$|\zrapstar|$}     & \multicolumn{1}{c}{\rfb} & \multicolumn{1}{c}{\fsrcorr (forward)} & \multicolumn{1}{c}{\fsrcorr (backward)} \\ \hline
$ [2.5,3.0]$     &       $ 0.75\pm 0.13\pm 0.27\pm 0.26$     &      $ 1.023\pm0.001$  &  $ 1.020\pm0.001$ \\
$ [3.0,3.5]$     &       $ 0.67\pm 0.12\pm 0.24\pm 0.24$     &      $ 1.030\pm0.001$  &  $ 1.018\pm0.001$ \\
$ [3.5,4.0]$     &       $ 0.89\pm 0.36\pm 0.71\pm 0.71$     &      $ 1.014\pm0.002$  &  $ 1.028\pm0.002$ \\
\hline\hline
\end{tabular}\vspace*{0.5cm}
\label{tab:diffrfb-y}
\end{center}
\end{table}

\begin{table}[htbp]
\begin{center}
\caption{The forward-backward ratio (\rfb) in intervals of \ZpT, together with the FSR correction.
For the \rfb results, the first uncertainty is statistical, the second is systematic and the third is from integrated luminosity.}
\begin{tabular}{rrccc}
\hline\hline
\multicolumn{1}{c}{\ZpT\,\unitgev}     & \multicolumn{1}{c}{\rfb} & \multicolumn{1}{c}{\fsrcorr (forward)} & \multicolumn{1}{c}{\fsrcorr (backward)} \\ \hline
$ [0.0,4.6]$        &       $ 0.67\pm 0.15\pm 0.04\pm 0.02$     &      $ 1.068\pm0.002$ &       $ 1.082\pm0.002$ \\
$ [4.6,10.5]$       &       $ 0.92\pm 0.27\pm 0.06\pm 0.03$     &      $ 1.033\pm0.001$ &       $ 1.025\pm0.002$ \\
$ [10.5,19.0]$      &       $ 1.25\pm 0.37\pm 0.08\pm 0.05$     &      $ 0.994\pm0.002$ &       $ 0.976\pm0.002$ \\
$ [19.0,34.0]$      &       $ 0.44\pm 0.14\pm 0.03\pm 0.02$     &      $ 0.988\pm0.002$ &       $ 0.981\pm0.002$ \\
$ [34.0,63.0]$      &       $ 1.03\pm 0.46\pm 0.08\pm 0.04$     &      $ 1.034\pm0.002$ &       $ 1.028\pm0.003$ \\
$ [63.0,270.0]$     &       $ 0.74\pm 0.59\pm 0.09\pm 0.03$     &      $ 1.043\pm0.007$ &       $ 1.064\pm0.009$ \\
\hline\hline
\end{tabular}\vspace*{0.5cm}
\label{tab:diffrfb-zpt}
\end{center}
\end{table}

\begin{table}[htbp]
\begin{center}
\caption{The forward-backward ratio (\rfb) in intervals of \phistar, together with the FSR correction.
For the \rfb results, the first uncertainty is statistical, the second is systematic and the third is from integrated luminosity.}
\begin{tabular}{rrccc}
\hline\hline
\multicolumn{1}{c}{\phistar}     & \multicolumn{1}{c}{\rfb} & \multicolumn{1}{c}{\fsrcorr (forward)} & \multicolumn{1}{c}{\fsrcorr (backward)} \\ \hline
$ [0.00, 0.02]$     &       $ 0.71\pm 0.16\pm 0.05\pm 0.03$     &      $ 1.031\pm0.003$ &       $ 1.053\pm0.003$ \\
$ [0.02, 0.05]$     &       $ 0.82\pm 0.19\pm 0.05\pm 0.03$     &      $ 1.044\pm0.002$ &       $ 1.007\pm0.002$ \\
$ [0.05, 0.10]$     &       $ 0.86\pm 0.22\pm 0.06\pm 0.03$     &      $ 1.010\pm0.002$ &       $ 1.015\pm0.002$ \\
$ [0.10, 0.20]$     &       $ 0.75\pm 0.19\pm 0.05\pm 0.03$     &      $ 1.018\pm0.001$ &       $ 1.014\pm0.001$ \\
$ [0.20, 0.40]$     &       $ 0.65\pm 0.21\pm 0.05\pm 0.02$     &      $ 1.004\pm0.002$ &       $ 1.015\pm0.002$ \\
$ [0.40, 4.00]$     &       $ 0.94\pm 0.46\pm 0.08\pm 0.03$     &      $ 1.024\pm0.004$ &       $ 1.023\pm0.004$ \\
\hline\hline
\end{tabular}\vspace*{0.5cm}
\label{tab:diffrfb-zps}
\end{center}
\end{table}

\begin{table}[htbp]
\begin{center}
\caption{The nuclear modification factors (\rpa) in intervals of \zrapstar for \forw and \backw collisions, together with the FSR correction.
For the \rpa results, the first uncertainty is statistical, the second is systematic and the third is from integrated luminosity.}
\begin{tabular}{rrrr}
\hline\hline
&\multicolumn{1}{c}{\zrapstar}    &     \multicolumn{1}{c}{\rpa}     & \multicolumn{1}{c}{\fsrcorr} \\ \hline
\For 
&$ [2.0,2.5]$       &       $ 0.87\pm 0.10\pm 0.03\pm 0.02$     &      $ 1.020\pm0.002$  \\
&$ [2.5,3.0]$       &       $ 0.98\pm 0.10\pm 0.04\pm 0.03$     &      $ 1.023\pm0.001$  \\
&$ [3.0,3.5]$       &       $ 0.89\pm 0.13\pm 0.03\pm 0.02$     &      $ 1.030\pm0.001$  \\
&$ [3.5,4.0]$       &       $ 0.83\pm 0.29\pm 0.03\pm 0.02$     &      $ 1.014\pm0.002$  \\
\hline
\Back 
&$ [-4.0,-3.5]$     &       $ 0.92\pm 0.17\pm 0.03\pm 0.03$     &      $ 1.028\pm0.002$  \\
&$ [-3.5,-3.0]$     &       $ 1.31\pm 0.14\pm 0.05\pm 0.05$     &      $ 1.018\pm0.001$  \\
&$ [-3.0,-2.5]$     &       $ 1.29\pm 0.19\pm 0.05\pm 0.05$     &      $ 1.020\pm0.001$  \\
\hline\hline
\end{tabular}\vspace*{0.5cm}
\label{tab:diffrpa-y}
\end{center}
\end{table}

\begin{table}[htbp]
\begin{center}
\caption{The nuclear modification factors (\rpa) in intervals of \ZpT for \forw and \backw collisions, together with the FSR correction.
For the \rpa results, the first uncertainty is statistical, the second is systematic and the third is from integrated luminosity.}
\begin{tabular}{rrrr}
\hline\hline
&\multicolumn{1}{c}{\ZpT\,\unitgev}    &     \multicolumn{1}{c}{\rpa}     & \multicolumn{1}{c}{\fsrcorr} \\ \hline
\For 
&$ [0.0,4.6]$        &       $ 1.04\pm 0.15\pm 0.04\pm 0.03$     &      $ 1.068\pm0.002$  \\
&$ [4.6,10.5]$       &       $ 0.77\pm 0.11\pm 0.03\pm 0.02$     &      $ 1.033\pm0.001$  \\
&$ [10.5,19.0]$      &       $ 0.83\pm 0.13\pm 0.03\pm 0.02$     &      $ 0.994\pm0.002$  \\
&$ [19.0,34.0]$      &       $ 0.89\pm 0.15\pm 0.03\pm 0.02$     &      $ 0.988\pm0.002$  \\
&$ [34.0,63.0]$      &       $ 1.34\pm 0.25\pm 0.06\pm 0.04$     &      $ 1.034\pm0.002$  \\
&$ [63.0,270.0]$     &       $ 1.41\pm 0.44\pm 0.10\pm 0.04$     &      $ 1.043\pm0.007$  \\
\hline
\Back 
&$ [0.0,4.6]$        &       $ 1.63\pm 0.24\pm 0.07\pm 0.04$     &      $ 1.082\pm0.002$  \\
&$ [4.6,10.5]$       &       $ 0.76\pm 0.17\pm 0.03\pm 0.02$     &      $ 1.025\pm0.002$  \\
&$ [10.5,19.0]$      &       $ 0.79\pm 0.18\pm 0.03\pm 0.02$     &      $ 0.976\pm0.002$  \\
&$ [19.0,34.0]$      &       $ 1.94\pm 0.39\pm 0.09\pm 0.05$     &      $ 0.981\pm0.002$  \\
&$ [34.0,63.0]$      &       $ 1.13\pm 0.40\pm 0.06\pm 0.03$     &      $ 1.028\pm0.003$  \\
&$ [63.0,270.0]$     &       $ 1.63\pm 0.98\pm 0.11\pm 0.04$     &      $ 1.064\pm0.009$  \\
\hline\hline
\end{tabular}\vspace*{0.5cm}
\label{tab:diffrpa-zpt}
\end{center}
\end{table}

\begin{table}[htbp]
\begin{center}
\caption{The nuclear modification factors (\rpa) in intervals of \phistar for \forw and \backw collisions, together with the FSR correction.
For the \rpa results, the first uncertainty is statistical, the second is systematic and the third is from integrated luminosity.}
\begin{tabular}{rrrr}
\hline\hline
&\multicolumn{1}{c}{\phistar}    &     \multicolumn{1}{c}{\rpa}     & \multicolumn{1}{c}{\fsrcorr} \\ \hline
\For 
&$ [0.00,0.02]$     &       $ 0.94\pm 0.13\pm 0.04\pm 0.02$     &      $ 1.031\pm0.003$  \\
&$ [0.02,0.05]$     &       $ 0.82\pm 0.11\pm 0.03\pm 0.02$     &      $ 1.044\pm0.002$  \\
&$ [0.05,0.10]$     &       $ 0.83\pm 0.11\pm 0.03\pm 0.02$     &      $ 1.010\pm0.002$  \\
&$ [0.10,0.20]$     &       $ 0.87\pm 0.13\pm 0.04\pm 0.02$     &      $ 1.018\pm0.001$  \\
&$ [0.20,0.40]$     &       $ 1.07\pm 0.18\pm 0.05\pm 0.03$     &      $ 1.004\pm0.002$  \\
&$ [0.40,4.00]$     &       $ 1.52\pm 0.29\pm 0.09\pm 0.04$     &      $ 1.024\pm0.004$  \\
\hline
\Back 
&$ [0.00,0.02]$     &       $ 1.32\pm 0.20\pm 0.05\pm 0.03$     &      $ 1.053\pm0.003$  \\
&$ [0.02,0.05]$     &       $ 1.07\pm 0.17\pm 0.04\pm 0.03$     &      $ 1.007\pm0.002$  \\
&$ [0.05,0.10]$     &       $ 0.94\pm 0.17\pm 0.04\pm 0.02$     &      $ 1.015\pm0.002$  \\
&$ [0.10,0.20]$     &       $ 1.22\pm 0.22\pm 0.05\pm 0.03$     &      $ 1.014\pm0.001$  \\
&$ [0.20,0.40]$     &       $ 1.57\pm 0.34\pm 0.07\pm 0.04$     &      $ 1.015\pm0.002$  \\
&$ [0.40,4.00]$     &       $ 1.34\pm 0.51\pm 0.08\pm 0.03$     &      $ 1.023\pm0.004$  \\
\hline\hline
\end{tabular}\vspace*{0.5cm}
\label{tab:diffrpa-zps}
\end{center}
\end{table}

\clearpage

\addcontentsline{toc}{section}{References}

\bibliographystyle{LHCb}
\bibliography{main,standard,LHCb-PAPER,LHCb-CONF,LHCb-DP,LHCb-TDR}

\newpage
\centerline
{\large\bf LHCb collaboration}
\begin
{flushleft}
\small
R.~Aaij$^{32}$\lhcborcid{0000-0003-0533-1952},
A.S.W.~Abdelmotteleb$^{50}$\lhcborcid{0000-0001-7905-0542},
C.~Abellan~Beteta$^{44}$,
F.~Abudin{\'e}n$^{50}$\lhcborcid{0000-0002-6737-3528},
T.~Ackernley$^{54}$\lhcborcid{0000-0002-5951-3498},
B.~Adeva$^{40}$\lhcborcid{0000-0001-9756-3712},
M.~Adinolfi$^{48}$\lhcborcid{0000-0002-1326-1264},
H.~Afsharnia$^{9}$,
C.~Agapopoulou$^{13}$\lhcborcid{0000-0002-2368-0147},
C.A.~Aidala$^{76}$\lhcborcid{0000-0001-9540-4988}
% ...

{\footnotesize \it

% ...
$^{9}$Universit{\'e} Clermont Auvergne, CNRS/IN2P3, LPC, Clermont-Ferrand, France\\
% ...
$^{13}$LPNHE, Sorbonne Universit{\'e}, Paris Diderot Sorbonne Paris Cit{\'e}, CNRS/IN2P3, Paris, France\\
% ...
$^{32}$Nikhef National Institute for Subatomic Physics, Amsterdam, Netherlands\\
% ...
$^{40}$Instituto Galego de F{\'\i}sica de Altas Enerx{\'\i}as (IGFAE), Universidade de Santiago de Compostela, Santiago de Compostela, Spain\\
% ...
$^{44}$Physik-Institut, Universit{\"a}t Z{\"u}rich, Z{\"u}rich, Switzerland\\
% ...
$^{48}$H.H. Wills Physics Laboratory, University of Bristol, Bristol, United Kingdom\\
% ...
$^{50}$Department of Physics, University of Warwick, Coventry, United Kingdom\\
% ...
$^{54}$Oliver Lodge Laboratory, University of Liverpool, Liverpool, United Kingdom\\
% ...
$^{62}$Syracuse University, Syracuse, NY, United States\\
% ...
$^{76}$University of Michigan, Ann Arbor, MI, United States, associated to $^{62}$\\
\bigskip
% ...
}
\end{flushleft}

\end{document}